\documentstyle[prb,preprint,eqsecnum,aps]{revtex}
\tighten
\begin{document}
\draft
\title{Order parameter description of the Anderson-Mott transition}

\author{D.Belitz}
\address{Department of Physics and Materials Science Institute,\\
University of Oregon,\\
Eugene, OR 97403}
\author{T.R.Kirkpatrick}
\address{Institute for Physical Science and Technology,\\
University of Maryland,\\
College Park, MD 20742}

\date{\today}
\maketitle

\begin{abstract}
An order parameter description of the Anderson-Mott transition (AMT) is
given. We first derive an order parameter field theory for the AMT, and
then present a mean-field solution. It is shown that the mean-field
critical exponents are exact above the upper critical dimension.
Renormalization group methods are then used to show that a random-field
like term is generated under renormalization. This leads to similarities
between the AMT and random-field magnets, and to an upper critical
dimension $d_{c}^{+}=6$ for the AMT. For $d<6$, an $\epsilon = 6-d$
expansion is used to calculate the critical exponents. To first order
in $\epsilon$ they are found to coincide with
the exponents for the random-field Ising model.
We then discuss a general scaling theory for the AMT. Some
well established scaling relations, such as Wegner's scaling law, are
found to be modified due to random-field effects.
New experiments are proposed to test for random-field
aspects of the AMT.
\end{abstract}
\narrowtext
\section{INTRODUCTION}
\label{sec:I}

Metal-insulator transitions of purely electronic origin, i.e. those
for which the structure of the ionic background does not play an
essential role, are commonly divided into two categories. In one
category the transition is triggered by electronic correlations, in
the other it is driven by disorder.\cite{Mott} The first case is known as a
Mott transition, and the second one as an Anderson transition. It
is believed that for many real metal-insulator transitions both
correlations and disorder are relevant, and this expectation is also
borne out by model studies of disordered interacting electron systems.\cite{R}
The resulting quantum phase transition, which carries aspects of both
types of transitions, we call an Anderson-Mott transition (AMT).

\par
Most of the theoretical work to date on the AMT has been based on a
small disorder expansion near $d=2$, which is believed to be the
lower critical dimension for the AMT.\cite{R} A generalized matrix nonlinear
$\sigma$-model (NL$\sigma$M) has proven very useful for this purpose.\cite{F1}
This model is a generalization of Wegner's theory for the Anderson
transition in noninteracting electronic systems.\cite{Wegner79}
It has led to the
identification of various universality classes for the AMT, as well as
to an $\epsilon = d-2$ expansion for the critical exponents.\cite{F2,Cetal}
Complementary work has recently been done by
adding disorder to models which display a Mott transition. In the limit
of infinite dimensions the resulting models have been studied in
some detail.\cite{Hubbardstuff}%

\par
{}From a standard phase transition point of view these approaches to the
AMT are quite unconventional. The standard procedure for solving phase
transition problems \cite{MaFisher}
is to (1) identify the relevant order parameter (OP),
and possibly derive or postulate an effective field theory for the OP,
(2) construct a Landau, or mean-field description of the phase transition,
and (3) use renormalization group (RG) methods to identify the upper
critical dimension, $d_{c}^{+}$, and to compute the critical exponents
in an $\epsilon = d_{c}^{+}-d$ expansion. RG methods can also be used
to derive a general scaling description of the phase transition for
$d<d_{c}^{+}$.

\par
Early attempts to implement this standard program for the Anderson
transition \cite{HarrisLubensky}
failed, because the most obvious simple OP, viz. the
single-particle density of states (DOS) at the Fermi
level,\cite{OPfootnote} turned
out to be uncritical.\cite{WegnerDOS}
As a result, the Anderson transition can only
be described by the NL$\sigma$M approach, or
possibly in terms of a complicated
functional OP.\cite{FyodorovMirlin}
The situation is, however, fundamentally different at
the AMT, where the DOS is generally believed to be critical, and to
vanish at the transition. This raises the prospect of using the DOS
for a conventional OP description of the AMT, making the AMT
conceptually simpler than the Anderson transition.

\par
In two previous short publications \cite{Letter1,Letter2}
we have shown that these expectations
are indeed justified, and have discussed such an OP approach to the AMT.
The purpose of the present paper is to explain the technical details of
the OP approach, and to present a number of additional results.

\par
One of the most far-reaching implications of our approach
is that the AMT is in some respects similar to magnetic transitions
in random fields.\cite{Letter2}
We will present the technical reasons underlying this
conclusion in Sec.\ \ref{sec:IV} below. Here we argue that a random-field
structure of the theory could have been anticipated on general physical
grounds. Let us consider a model of an interacting disordered electron
gas. In terms of anticommuting Grassmann fields, $\bar\psi$ and $\psi$,
the action can be written,\cite{R}
\begin{mathletters}
\label{eqs:1.1}
\begin{eqnarray}
S=-\sum_{\sigma} \int dx\ \bar\psi_{\sigma}(x)\biggl[{\partial\over \partial
   \tau} - {1\over 2m}\nabla ^{2} - \mu + u({\bf x})\biggr]\psi_{\sigma}(x)
\nonumber\\
  -{\Gamma\over 2}\ \sum_{\sigma_{1}\sigma_{2}}\int dx\
                                               \bar\psi_{\sigma_{1}}(x)
   \bar\psi_{\sigma_{2}}(x)\psi_{\sigma_{2}}(x)\psi_{\sigma_{1}}(x)
   \quad.
\label{eq:1.1a}
\end{eqnarray}
Here $x\equiv ({\bf x},\tau)$ with $\tau$ denoting imaginary time,
$\int dx \equiv \int d{\bf x} \int_{0}^{1/T} d\tau$, $m$ is the electron mass,
$\mu$ is the chemical potential, $\sigma$ is a spin label, and for
simplicity we have assumed an instantaneous point-like electron-electron
interaction with strength $\Gamma$. $u({\bf x})$ is a random potential
which represents the disorder. We assume $u$ to be $\delta$-correlated,
and to obey a Gaussian distribution with second moment
\begin{equation}
\{u({\bf x})u({\bf y})\} = {1\over 2\pi N_{F}\tau_{el}}\
\delta({\bf x}-{\bf y}) \quad,
\label{eq:1.1b}
\end{equation}
\end{mathletters}%
where the braces denote the disorder average, $N_{F}$ is the bare DOS
per spin at the Fermi energy, and $\tau_{el}$ is the bare elastic mean-free
time. For future reference we denote the disorder related part of the
action by
\begin{equation}
S_{dis}=-\sum_{\sigma}\int dx\ u({\bf x})\
                       \bar\psi_{\sigma}(x) \psi_{\sigma}(x)
       = -\sum_{\sigma,n}\int d{\bf x}\ u({\bf x})\
                                  \bar\psi_{\sigma,n}({\bf x})
           \psi_{\sigma,n}({\bf x})\quad.
\label{eq:1.2}
\end{equation}
In the second equality in Eq.\ (\ref{eq:1.2}) a Matsubara frequency
decomposition of $\bar\psi(\tau)$ and $\psi(\tau)$ has been used.

\par
As mentioned above, the most obvious OP for the AMT is the single-particle
DOS, $N$, at the Fermi level. In terms of Grassmann variables this quantity
is proportional to the zero-frequency limit of the expectation value
of the composite fermionic variable $\bar\psi\psi$: $N={\rm Im} N(i\omega_{n}
\rightarrow 0+i0)$, with
\begin{equation}
N(i\omega_{n})={-1\over 2\pi N_{F}}\sum_{\sigma}\
               \langle\bar\psi_{\sigma,n}({\bf x})
               \psi_{\sigma,n}({\bf x})\rangle\quad,
\label{eq:1.3}
\end{equation}
where we have normalized the DOS by $2N_{F}$.
Equations (\ref{eq:1.2},\ \ref{eq:1.3}) suggest that the OP for the AMT couples
directly to the random potential $u$, and that this random field (RF) term
is structurally identical to the one that appears in magnetic RF
problems.\cite{ImryMa,Grinstein}
Notice that this structure appears in both interacting and noninteracting
localization theories. However, in the interacting case there is an
additional physical feature:
The interaction will in general favor a local electron
arrangement that is different from the one favored by the random potential.
This kind of frustration is crucial for the generation of the typical RF
effects which are known from RF magnets. One should therefore expect RF
effects at the AMT, but not necessarily at the Anderson transition.

\par
This conclusion has several important implications. For instance, one
expects hyperscaling to be violated at the AMT due to a dangerous
irrelevant variable, as it is in RF magnets.\cite{Grinstein}
We will see below that this is indeed the case and that, as
a result, Wegner's scaling law \cite{Wegner76} relating
the conductivity exponent, $s$,
to the correlation length exponent, $\nu$, is modified.
\par
The plan of this paper is as follows. In Sec.\ \ref{sec:II} we
review the NL$\sigma$M approach to the AMT, and use it to derive an OP
field theory for the transition.
In Sec.\ \ref{sec:III} the OP field theory is solved in the saddle
point or mean-field approximation. We show that the mean-field solution
is stable for $d>d_{c}^{+}$, and that the mean-field exponents are
exact in this regime.
In Sec.\ \ref{sec:IV} RG methods are used to prove that the theory has
RF aspects, to establish that $d_{c}^{+}=6$, and to set up an
$\epsilon = d-6$ expansion for the critical exponents. In Sec.\ \ref{sec:V}
we discuss a general scaling description of the AMT. We conclude in
Sec.\ \ref{sec:VI} with a summary of the paper and a discussion of some
open problems as well as of experimental implications of our results.

\section{FORMALISM}
\label{sec:II}

In this section we start by briefly reviewing the field theoretic description
of disordered interacting electron systems. We focus on the NL$\sigma$M,
which is a free field theory with nonlinear constraints relating massive
and massless modes in the ordered phase, which in the case of the AMT
is the metallic one. We then integrate out the massless excitations to
obtain a field theory solely in terms of the massive modes, which are
directly related to the OP for the transition, i.e. the DOS.
This is different from the usual treatment of the NL$\sigma$M, which
integrates out the massive fluctuations to obtain a theory for the soft
modes. Our procedure is closely analogous to the treatment of the
$O(n)$ symmetric NL$\sigma$M in the limit of large $n$.\cite{ZJ}

\subsection{The Model}
\label{subsec:II.A}

We will start from the action given by Eq.\ (\ref{eq:1.1a}).
The NL$\sigma$M for the AMT can be derived from Eq.\ (\ref{eq:1.1a}) by
assuming that all of the relevant physics can be expressed in terms of
long-wavelength and low-frequency fluctuations of the number density,
the spin density, and the single-particle spectral density. Technically
this is achieved by making long-wavelength approximations, and by
introducing classical composite variables that are related to the
operators mentioned above. The quenched disorder is handled by means of
the replica trick. The resulting action reads,\cite{F1,R}
\begin{mathletters}
\label{eqs:2.1}
\begin{eqnarray}
S[\tilde Q]=&&-{1\over{2G}}\int d{\bf x}\;tr{\Bigl(}\nabla
               \tilde Q({\bf x}){\Bigr)}^2
     + 2H\int d{\bf x}\;tr{\Bigl(}\Omega \tilde Q({\bf x}){\Bigr)}
\nonumber\\
     &&-{\pi T \over 4}\sum_{u=s,t} \int d{\bf x}\
     \bigl[ \tilde Q({\bf x}) \gamma^{(u)} \tilde Q({\bf x})\bigr]\quad,
\label{eq:2.1a}
\end{eqnarray}
where
\begin{eqnarray}
\bigl[ \tilde Q({\bf x}) \gamma^{(s)} \tilde Q({\bf x})\bigr]=
     K_{s}\ \sum_{n_{1},n_{2},
     n_{3},n_{4}}\delta_{n_{1}+n_{3},n_{2}+n_{4}}\sum_{\alpha}\sum_{r=0,3}
     (-)^{r}\nonumber\\
\times
tr \Bigl( (\tau_{r}\otimes s_{0})\tilde Q_{n_{1}n_{2}}^{\alpha \alpha}\Bigr)
tr \Bigl( (\tau_{r}\otimes s_{0})\tilde Q_{n_{3}n_{4}}^{\alpha \alpha}\Bigr)
\quad,
\label{eq:2.1b}
\end{eqnarray}
and
\begin{eqnarray}
\bigl[ \tilde Q({\bf x}) \gamma^{(t)} \tilde Q({\bf x})\bigr]
     =-K_{t}\ \sum_{n_{1},n_{2},
     n_{3},n_{4}}\delta_{n_{1}+n_{3},n_{2}+n_{4}}\sum_{\alpha}\sum_{r=0,3}
     (-)^{r}\sum_{i=1}^{3}\nonumber\\
\times
tr \Bigl( (\tau_{r}\otimes s_{i})\tilde Q_{n_{1}n_{2}}^{\alpha \alpha}\Bigr)
tr \Bigl( (\tau_{r}\otimes s_{i})\tilde Q_{n_{3}n_{4}}^{\alpha \alpha}\Bigr)
\quad,
\label{eq:2.1c}
\end{eqnarray}
\end{mathletters}%
Here $\tilde Q$ is a classical field that is, roughly speaking,
composed of two fermionic fields. It carries two Matsubara frequency
labels, $n$ and $m$, and two replica labels, $\alpha$ and $\beta$.
The matrix elements $\tilde Q_{nm}^{\alpha \beta}$ are spin quaternions,
with the quaternion degrees of freedom describing the particle-hole
($\tilde Q \sim \bar\psi \psi$)
and particle-particle ($\tilde Q \sim \bar\psi\bar\psi$) channels,
respectively. We will restrict ourselves
to the particle-hole degrees of freedom.
The matrix elements can then be expanded in
a restricted spin-quaternion basis,
\begin{equation}
\tilde Q_{nm}^{\alpha \beta} = \sum_{r=0,3}\ \sum_{i=0}^{3}\ {_{r}^{i}\tilde Q
 _{nm}^{\alpha\beta}}\ (\tau_{r}\otimes s_{i})\quad,
\label{eq:2.2}
\end{equation}
with $\tau_{0,1,2,3}$ the quaternion basis, and $s_{0,1,2,3}$ the spin
basis ($s_{1,2,3}$ are the Pauli matrices). The matrix $\tilde Q$ is
subject to the nonlinear constraints,
\begin{mathletters}
\label{eqs:2.3}
\begin{equation}
\tilde Q^{2} = 1\quad,
\label{eq:2.3a}
\end{equation}
\begin{equation}
tr\ \tilde Q = 0\quad,
\label{eq:2.3b}
\end{equation}
\begin{equation}
\tilde Q^{\dagger} = C^{T}\tilde Q^{T} C = \tilde Q\quad.
\label{eq:2.3c}
\end{equation}
\end{mathletters}%
The last equation expresses the requirements of hermiticity and charge
conjugation; the matrix $C$ is block-diagonal with elements $i\tau_{1}
\otimes s_{2}$. Equation (\ref{eq:2.3c}) implies,
\begin{mathletters}
\label{eqs:2.4}
\begin{equation}
{^{i}_{r}\tilde Q_{nm}^{\alpha\beta}}^{*} = {^{i}_{r}\tilde
                                                    Q_{nm}^{\alpha\beta}}
\quad,
\label{eq:2.4a}
\end{equation}
\begin{equation}
{^{i}_{r}\tilde Q_{nm}^{\alpha\beta}} = (-)^{r}\ S_{i}\
                           {^{i}_{r}\tilde Q_{nm}^{\alpha\beta}}\quad,
\label{eq:2.4b}
\end{equation}
\end{mathletters}%
with $S_{0}=1$ and $S_{1,2,3}=-1$.

\par
In Eqs.\ (\ref{eqs:2.1}), $G=2/\pi\sigma$,
with $\sigma$ the bare conductivity,
is a measure of the disorder, and $H=\pi N_{F}/2$
is a frequency coupling parameter. $K_{s}$ and
$K_{t}$ are bare interaction amplitudes in the spin singlet and spin
triplet channels, respectively, and $\Omega_{nm}^{\alpha\beta}=
\delta_{nm}\delta_{\alpha\beta}\ (\tau_{0}\otimes s_{0})\ \omega_{n}$, with
$\omega_{n} = 2\pi Tn$, is a bosonic frequency matrix. Notice that
$K_s<0$ for repulsive interactions.

\par
The correlation functions of the $\tilde Q$ determine the physical
quantities. Correlations of $\tilde Q_{nm}$ with $nm<0$ determine the
diffusive modes which describe charge, spin, and heat diffusion, while
the DOS is determined by $\langle\tilde Q_{nn}^{\alpha\alpha}\rangle$,
cf. Eq.\ (\ref{eq:3.3a}) below. It is therefore convenient to separate
$\tilde Q$ into blocks,
\begin{eqnarray}
\tilde{Q}^{\alpha\beta}_{nm}=
\Theta(nm)Q^{\alpha\beta}_{nm}({\bf x})+\Theta(n)\Theta(-
m)q^{\alpha\beta}_{nm}({\bf x})
\nonumber\\
+\Theta(-n)\Theta(m)(q^{\dagger})^{\alpha\beta}_{nm}({\bf x})\quad.
\label{eq:2.5}
\end{eqnarray}
Normally a NL$\sigma$M is treated by integrating out the massive modes,
i.e. the $Q_{nm}$ in this case, to obtain an effective
theory for the massless modes, which here are the diffusion processes
described by $q$ and $q^\dagger$.
However, since our goal is to obtain a field theory for the OP for the AMT,
$Q_{nn}$, we will proceed differently, and instead integrate out the
massless $q$-fields. To this end, we use a functional integral representation
of the delta function,
\begin{equation}
\prod_{\bf x}\ \delta\bigl[\tilde Q^{2}({\bf x})-1\bigr] =
 \int D[\Lambda]\ \exp\Bigl\{-{1\over 2G}\int d{\bf x}\ tr\ \bigl(\Lambda
 ({\bf x})[\tilde Q^{2}({\bf x})-1]\bigr)\Bigr\}\quad,
\label{eq:2.6}
\end{equation}
where $tr$ denotes a trace over all discrete indices, and
the factor of $1/2G$ has been inserted for convenience. Together
with Eq.\ (\ref{eq:2.1a}) this allows us to write the action as,
\begin{eqnarray}
S_{1}[\tilde{Q},\Lambda]=\frac{-1}{2G}\int d{\bf x}\ tr
\biggl[\Lambda({\bf x})[\tilde{Q}^2({\bf x})-
\openone]+\Bigl(\partial_{\bf x}\tilde{Q}({\bf x})\Bigr)^2\biggr]+2H \int
d{\bf x}\ tr\Bigl(\Omega \tilde{Q}({\bf x})\Bigr)
\nonumber\\
-\frac{\pi T}{4} \sum_{u=s,t} [\tilde{Q}({\bf x})\gamma^{(u)} \tilde{Q}({\bf
x})]\quad.
\label{eq:2.7}
\end{eqnarray}
We note that the constraint given by Eq.\ (\ref{eq:2.3b}) does not involve
$q$, so it can be imposed later.\cite{tracefootnote}
Furthermore, the contraints expressed
by Eqs.\ (\ref{eq:2.3c}) or (\ref{eqs:2.4}) do not couple independent blocks
of $\tilde Q$. Finally, by decomposing $\Lambda$ into blocks like
$\tilde Q$ one sees that the elements $\Lambda_{nm}$ with $nm>0$,
together with the tracelessness condition, Eq.\ (\ref{eq:2.3b}), are
sufficient for enforcing the constraint $\tilde Q^{2}=1$. We can
therefore restrict ourselves to $\Lambda_{nm}$ with $nm>0$.

\par
The partition function of the replicated theory (N replicas) is given by,
\begin{eqnarray}
Z^{N} = \int D[\tilde Q]\ \exp\bigl(S[\tilde Q]\bigr)\nonumber\\
      = \int D[\tilde Q]\ D[\Lambda]\ \exp\bigl(S_{1}[\tilde Q,\Lambda]\bigr)
      = \int D[Q]\ D[\Lambda]\ \exp\bigl(S_{2}[Q,\Lambda]\bigr)\quad.
\label{eq:2.8}
\end{eqnarray}
In the last equality in Eq.\ (\ref{eq:2.8}) we have defined
yet another action,
\begin{equation}
S_{2}[Q,\Lambda] = \ln\ \int D[q,q^{\dagger}]\ \exp\bigl(S_{1}[\tilde Q,
 \Lambda]\bigr)\quad,
\label{eq:2.9}
\end{equation}
by integrating out the massless $q$-fields.
Since the action $S_{1}$, Eq.\ (\ref{eq:2.7}), is quadratic in these
fields, this can be done exactly. Formally one obtains,
\begin{mathletters}
\label{eqs:2.10}
\begin{equation}
S_{2}[Q,\Lambda] = S_{1}[Q,\Lambda] - {1\over 2}\ Tr\ \ln M[\Lambda]\quad.
\label{eq:2.10a}
\end{equation}
Here $Tr$ denotes a trace over {\it all} degrees of freedom, and
$M$ is a matrix with elements,
\begin{eqnarray}
{_{r_{1}r_{2}}^{i_{1}i_{2}}M_{n_{1}n_{2},n_{3}n_{4}}^{\alpha_{1}\alpha_{2},
                                      \alpha_{3}\alpha_{4}}}\ ({\bf x}) =
{\delta_{\alpha_{2}\alpha_{4}}\delta_{n_{2}n_{4}}\over 2G}\ \sum_{r_{3},
          i_{3}}\ tr\ \bigl(\tau_{r_{3}}\tau_{r_{1}}\tau_{r_{2}}^{\dagger}
          \bigr)\ tr\ \bigl(s_{i_{3}}s_{i_{1}}s_{i_{2}}^{\dagger}\bigr)\
          {_{r_{3}}^{i_{3}}\Lambda_{n_{3}n_{1}}^{\alpha_{3}\alpha_{1}}}
          ({\bf x})\nonumber\\
+{\delta_{\alpha_{1}\alpha_{3}}\delta_{n_{1}n_{3}}\over 2G}\ \sum_{r_{3},
          i_{3}}\ tr\ \bigl(\tau_{r_{1}}\tau_{r_{3}}\tau_{r_{2}}^{\dagger}
          \bigr)\ tr\ \bigl(s_{i_{1}}s_{i_{3}}s_{i_{2}}^{\dagger}\bigr)\
          {_{r_{3}}^{i_{3}}\Lambda_{n_{2}n_{4}}^{\alpha_{2}\alpha_{4}}}
          ({\bf x})\nonumber\\
+ {4\over G}\ \delta_{\alpha_{1}\alpha_{3}}\delta_{\alpha_{2}\alpha_{4}}
          \delta_{r_{1}r_{2}}\delta_{i_{1}i_{2}}\Bigl[\delta_{n_{1}n_{3}}
          \delta_{n_{2}n_{4}}\bigl(-\nabla^{2}\bigr) + \delta_{\alpha_{1}
          \alpha_{2}}\delta_{n_{1}+n_{4},n_{2}+n_{3}} 2\pi TGK_{\nu_{i_{1}}}
          \Bigr]\quad,
\label{eq:2.10b}
\end{eqnarray}
\end{mathletters}%
with $\nu_{0}=s$ and $\nu_{1,2,3}=t$. In giving Eq.\ (\ref{eq:2.10a}) we have
omitted some terms which result from the interaction, i.e. the last term
in Eq.\ (\ref{eq:2.7}), coupling $Q$ and $q$. These terms will turn out
to be irrelevant (in the RG sense)
because they are of higher order in the frequency.

\subsection{The Order Parameter Field Theory}
\label{subsec:II.B}

The field theory given by Eqs.\ (\ref{eqs:2.10}) has been formulated in terms
of the OP for the AMT and the auxiliary, or ghost, field $\Lambda$. In order
to obtain a theory solely in terms of $Q$ we need to integrate out $\Lambda$.
We will do this perturbatively, and verify later that the neglected terms are
either RG irrelevant, or provide analytic corrections to nonuniversal
coefficients in the theory. Either way they do not modify the critical
behavior.
Physically this is to be expected, since it is easily established that the
$\Lambda$-modes are not critical near the AMT.\cite{nonpertfootnote}

\par
To proceed we write the average of $\Lambda$ as
\begin{mathletters}
\label{eqs:2.11}
\begin{equation}
\langle {_{r}^{i}\Lambda_{nm}^{\alpha\beta}}({\bf x})\rangle =
       \delta_{r0}\delta_{i0}\delta_{nm}\ell_n\quad,
\label{eq:2.11a}
\end{equation}
with $\ell_{n}$ to be determined later. We write
\begin{equation}
{_{r}^{i}\Lambda_{nm}^{\alpha\beta}}({\bf x}) =
     \langle {_{r}^{i}\Lambda_{nm}^{\alpha\beta}}({\bf x})\rangle +
     {_{r}^{i}\psi_{nm}^{\alpha\beta}}({\bf x})\quad,
\label{eq:2.11b}
\end{equation}
\end{mathletters}%
and expand in powers of $\psi$ to determine a new OP action, $S_{3}[Q]$, which
is defined by
\begin{equation}
S_{3}[Q] = \ln\ \int D[\psi]\ \exp \bigl(S_{2}[Q,\langle\Lambda\rangle + \psi]
             \bigr)\quad.
\label{eq:2.12}
\end{equation}
Integrating over $\psi$ in Gaussian approximation,
neglecting constant contributions to the action,
and neglecting explicit interaction terms (see the end of this subsection
for a justification), we obtain
\begin{eqnarray}
S_{3}[Q] = -{1\over 2G}\ \int d{\bf x}\ tr\ \Bigl[\bigl(\nabla Q({\bf x})\bigr)
           ^{2} + \langle\Lambda\rangle Q^{2}({\bf x})\Bigr] +
           2H\int d{\bf x}\ tr\ \bigl[\Omega Q({\bf x})\bigr]\nonumber\\
+ {u\over 2G^{2}}\ \int d{\bf x}\ tr\ \bigl[(1-f)Q^{2}({\bf x})\bigr] -
    {u\over 4G^{2}}\ \int d{\bf x}\ tr\ Q^{4}({\bf x})\nonumber\\
- {v\over 4G^{2}}\
    \int d{\bf x}\ \bigl(tr_{+}Q^{2}({\bf x})\bigr)\bigl(tr_{-}Q^{2}({\bf
x})\bigr)
    \quad,
\label{eq:2.13}
\end{eqnarray}
where $tr_{\pm}$ denotes 'half-traces' that sum over all
replica labels but only over positive and negative
frequencies, respectively: $tr_+ = \sum_{\alpha}\,\sum_{n\ge 0}$\ ,
$tr_- = \sum_{\alpha}\,\sum_{n<0}$\ . $f=f(\langle\Lambda\rangle)$
is a matrix with elements
${_{r}^{i}f_{nm}^{\alpha\beta}}({\bf x}) =
\delta_{r0}\delta_{i0}\delta_{nm}f_{n}$ with,
\begin{equation}
f_{n} = - \frac{G}{4}\int_{\bf p}\sum^{-{\infty}}_{m=-1}\frac{2\pi
TG}{[p^2+\frac{1}{2}(\ell_{n} +\ell_{m})]^2}\ \sum_{i=0}^{3}\ K_{\nu_{i}}
\left[1+\sum^{n-m-1}_{n_{1}=0}\frac{2 \pi TGK_{\nu_{i}}}{p^2 +
\frac{1}{2}(\ell_{n_{1}} + \ell_{n_{1} - n+m})}
\right]^{-1}\quad,
\label{eq:2.14}
\end{equation}
for $n\ge 0$, where $\int_{\bf p}=(2\pi)^{-d}\int d{\bf p}$. For $n\le 0$,
the same expression holds with the sum over $m$
from $0$ to $\infty$, and with $n$ and $m$ interchanged in the summand. The
coefficients $u$ and $v$ in Eq.\ (\ref{eq:2.13}) are given by,
\begin{mathletters}
\label{eqs:2.15}
\begin{equation}
u = -G/\bigl(df_{n}/d\ell_{n}\bigr)_{n=0}\quad,
\label{eq:2.15a}
\end{equation}
and
\begin{equation}
v = -u^{2}\ \int_{\bf p}\ {1\over p^{4}}\quad.
\label{eq:2.15b}
\end{equation}
\end{mathletters}%
In writing Eq.\ (\ref{eq:2.13}) we have localized the coefficients $u$ and
$v$ in space and imaginary time. Frequency and wavenumber dependent
corrections to this result turn out to either renormalize nonuniversal
coefficients, or to be RG irrelevant. Note that $v$ is infinite for
$d\le 4$. $u$ also contains singularities in $d\le 4$. In
the main text of this paper
we restrict ourselves to the region $d\approx d_{c}^{+}=6$, and ignore
these singularities. In Appendix \ref{app:A} we show that, at the mean-field
level, they are of no consequence for the critical behavior
as long as $d>3$.
Beyond the mean-field level, these singularities are one
of several severe difficulties one encounters. This point will be further
discussed in Sec.\ \ref{sec:VI}.
Finally, we note that our integrating out of $\Lambda$ has obscured the
condition
\begin{mathletters}
\label{eqs:2.16}
\begin{equation}
\langle\psi\rangle = 0\quad,
\label{eq:2.16a}
\end{equation}
which follows from Eq.\ (\ref{eq:2.11b}). By returning to the action $S_{2}$
one sees that Eq.\ (\ref{eq:2.16a}) implies,
\begin{equation}
0 = 1 - f_{n}(\langle\Lambda\rangle) - \sum_{\beta,m,r,i}\ (-)^{r}\ S_{i}\
     \langle{^{i}_{r}Q_{nm}^{\alpha\beta}({\bf x})}
            {^{i}_{r}Q_{mn}^{\beta\alpha}({\bf x})}\rangle + \ldots\quad,
\label{eq:2.16b}
\end{equation}
\end{mathletters}%
with $S_{i}$ given after Eq.\ (\ref{eq:2.4b}).
Corrections to Eq.\ (\ref{eq:2.16b}) come from higher orders in the
$\psi$-expansion. We will show below that they do not change the critical
behavior.

\par
We conclude this section by mentioning two approximations that were used to
derive Eqs.\ (\ref{eq:2.13}) and (\ref{eqs:2.16}). First, the coefficients
given in Eq.\ (\ref{eq:2.13}) follow by
expanding $S_{2}$ in Eq.\ (\ref{eq:2.12})
to $O(\psi^{2})$, neglecting all terms of higher order in $\psi$. Second,
all explicit interaction terms, such as the last terms
in Eq.\ (\ref{eq:2.1a}), or the terms mentioned after Eq.\ (\ref{eq:2.10b}),
have been neglected in Eq.\ (\ref{eq:2.13}).
In the next section we will use RG methods to argue that these approximations
are justified, at least in the vicinity of $d_{c}^{+}$.

\section{MEAN-FIELD THEORY}
\label{sec:III}

In this section we solve the field theory derived in Sec.\ \ref{sec:II} in the
mean-field or saddle point approximation. We will show that the mean-field
theory describes a phase transition, and that it is an AMT. We then show that
the saddle point solution is locally stable, and that the mean-field exponents
are exact for $d>d_{c}^{+}$.

\subsection{Mean-Field Solution}
\label{subsec:III.A}

We can use either Eqs.\ (\ref{eqs:2.10}) or
(\ref{eq:2.13}) - (\ref{eqs:2.16})
to construct a mean-field theory. We choose to use the latter. We look for
saddle point (SP) solutions $Q_{sp}$, $\Lambda_{sp}$
that are spatially uniform and satisfy
\begin{mathletters}
\label{eqs:3.1}
\begin{equation}
{_{r}^{i}(Q_{sp})_{nm}^{\alpha\beta}} =
\delta_{r0}\delta_{i0}\delta_{nm}\delta_{\alpha\beta}N_{n}^{(0)}\quad,
\label{eq:3.1a}
\end{equation}
\begin{equation}
{_{r}^{i}(\Lambda_{sp})_{nm}^{\alpha\beta}} =
\delta_{r0}\delta_{i0}\delta_{nm}\delta_{\alpha\beta}\ell_{n}^{(0)}\quad,
\label{eq:3.1b}
\end{equation}
\end{mathletters}%
where the superscript $(0)$ denotes the SP approximation.
The replica, frequency, and spin-quaternion structures in
Eqs.\ (\ref{eqs:3.1}) are motivated by the fact
that $\langle {^{i}_{r}Q_{nm}^{ij}}\rangle$ and
$\langle {^{i}_{r}\Lambda_{nm}^{ij}}\rangle$ have these properties (cf.
Eq.\ (\ref{eq:2.11a})), and that in the mean-field approximation averages are
replaced by the corresponding SP values.

\par
Equations (\ref{eqs:3.1}) and (\ref{eq:2.16b}) yield
\begin{mathletters}
\label{eqs:3.2}
\begin{equation}
\bigl(N_{n}^{(0)}\bigr)^{2} = 1 - f_{n}(\langle\Lambda\rangle)\quad.
\label{eq:3.2a}
\end {equation}
Taking the extremum of the action $S_{3}$, Eq.\ (\ref{eq:2.13}), with respect
to $Q$ or $N^{(0)}$ gives
\begin{equation}
\ell_{n}^{(0)} = 2GH\omega_{n}/N_{n}^{(0)}\quad,
\label{eq:3.2b}
\end{equation}
\end{mathletters}%
where we have used Eq.\ (\ref{eq:3.2a}).

\par
Let us discuss some aspects of these results. First, the DOS at a frequency
or energy $\Omega$, measured from the Fermi surface,
and normalized by $2N_F$ (cf. Eq.\ (\ref{eq:1.3})), is given
by,\cite{F1}
\begin{mathletters}
\label{eqs:3.3}
\begin{equation}
N(\Omega) = {\rm Re}\ \langle {^{0}_{0}Q_{nn}^{\alpha\alpha}}({\bf x})
            \rangle_{i\omega_{n}\rightarrow \Omega + i0}
          \equiv\ {\rm Re}\ N_{n}\vert_{i\omega_{n}\rightarrow
                            \Omega + i0}\quad.
\label{eq:3.3a}
\end{equation}
In mean-field approximation this yields
\begin{equation}
N(\Omega) = N_{n}^{(0)}\vert_{i\omega_{n}\rightarrow \Omega + i0}\quad.
\label{eq:3.3b}
\end{equation}
\end{mathletters}%
Clearly, the DOS decreases with increasing disorder $G$. Furthermore,
$N(\Omega)$ is nonanalytic at $\Omega =0$, see Appendix \ref{app:A}.
If we insert Eqs.\ (\ref{eqs:3.2})
into Eq.\ (\ref{eq:3.3b}), and iterate to first order in $G$, then we just
recover the well-known 'Coulomb anomaly' of the DOS.\cite{AA} With further
increasing disorder, $N^{(0)}(\Omega =0)$ eventually vanishes at a critical
value $G_{c}$ of $G$. The mean-field approximation thus describes a phase
transition with a vanishing DOS at the Fermi level, the hallmark of the AMT.
(Transport properties we will discuss shortly). Note that in the absence of
interactions the quantity $f$, Eq.\ (\ref{eq:2.14}), vanishes, and hence
$N^{(0)}\equiv 1$. This reflects the fact that the DOS does {\it not}
vanish at an Anderson transition.\cite{WegnerDOS}

\par
Second, it is easy to determine the behavior of $N^{(0)}$ in the vicinity
of $G_{c}$. This is true even though Eqs.\ (\ref{eqs:3.2}) and
(\ref{eq:2.14}) represent an integral equation for $N_{n}^{(0)}$. The
point is that $\ell_{m}$ in Eq.\ (\ref{eq:2.14}) gets integrated over
all frequencies. At nonzero $m$ it is given by $\omega_{m}$ times a
finite, nonuniversal prefactor which depends on $N_{m}^{(0)}$, see
Eq.\ (\ref{eq:3.2b}). This implies that we can treat $\ell_{m}$ as a
finite, nonuniversal quantity. The distance from the critical point,
$t$, is proportional to $G-G_{c}$. In mean-field approximation $t$ is
given by,
\begin{equation}
t^{(0)} = 1 - f_{n=0}\quad.
\label{eq:3.4}
\end{equation}
Near the critical point we have
\begin{equation}
N_{n=0}^{(0)} = \bigl(t^{(0)}\bigr)^{1/2}\quad,
\label{eq:3.5}
\end {equation}
which yields the mean-field value of the critical exponent $\beta$,
\begin{equation}
\beta = 1/2\quad.
\label{eq:3.6}
\end{equation}
Note that the sign of $N^{(0)}_{n=0}$ in Eq.\ (\ref{eq:3.5}) is determined
by the way the zero-frequency limit is taken: $N_{n}$ is an odd function
of $n$. In this respect the external Matsubara frequency acts like a
symmetry breaking external field at the mean-field AMT.
The DOS, of course, is positive definite since in order to obtain it
from $N_{n}$ the branch cut must
always be approached from above, Eqs.\ (\ref{eqs:3.3}). To put it another
way, $N_{n}$ is $i$ times a causal function that is an odd function of
complex frequency (viz. the Green function). $N(\Omega)$ is the spectrum
of that function, which is even in $\Omega$.

\par
Third, it is not much more difficult to obtain all other critical exponents.
Let us write $Q$ as its expectation value plus fluctuations, as we
did with $\Lambda$ in Eq.\ (\ref{eq:2.11b}),
\begin{equation}
{^{i}_{r}Q_{nm}^{\alpha\beta}} = \delta_{r0}\delta_{i0}
         \delta_{\alpha\beta}\delta_{nm}N_{n} + \sqrt{2G}\
         {^{i}_{r}\phi_{nm}^{\alpha\beta}}\quad,
\label{eq:3.7}
\end{equation}
where the factor of $\sqrt{2G}$ has been inserted for later convenience.
The action governing Gaussian fluctuations about the mean-field solution
in the critical region then follows from Eq.\ (\ref{eq:2.13}) as,
\begin{eqnarray}
S_{3}[\phi] = -\int d{\bf x}\ tr\ \Bigl[\bigl(\nabla\phi({\bf x})\bigr)^{2}
              + \ell^{(0)}\phi^{2}({\bf x})
              + \frac{2u}{G}\ t^{(0)}\phi^{2}({\bf x})\Bigr]\nonumber\\
-\frac{v}{2G}t^{(0)}\ \int d{\bf x}\ \bigl(tr_{+}\phi({\bf x})\bigr)
              \bigl(tr_{-}\phi({\bf x})\bigr) + O(\phi^{3})\quad.
\label{eq:3.8}
\end{eqnarray}
All remaining critical exponents can now be read off Eq.\ (\ref{eq:3.8}):
Comparing the first and the third terms on the r.h.s. yields the
correlation length exponent $\nu=1/2$. With $\ell^{(0)}\sim \omega/Q$,
the first and the second term give the dynamical exponent $z=3$. (This
result one can also confirm by explicitly calculating $N(\Omega
\rightarrow 0)$ at $G=G_{c}$ from the mean-field equation,
see Appendix \ref{app:A}). Finally,
inspection shows that the $\phi$-$\phi$ correlation function
near the transition in the limit
of small frequencies and small wavenumbers has a standard Ornstein-Zernike
form, which yields exponents $\gamma=1$ and $\eta=0$. We thus have
standard mean-field values for all static exponents,
\begin{mathletters}
\label{eqs:3.9}
\begin{equation}
\beta=\nu=1/2\quad,\quad\gamma=1\quad,\quad\eta=0\quad,\quad\delta=3\quad,
\label{eq:3.9a}
\end{equation}
and for the dynamical exponent we have,
\begin{equation}
z=3\quad.
\label{eq:3.9b}
\end{equation}
\end{mathletters}%
Inspection of Eq.\ (\ref{eq:3.8}) further shows that the AMT saddle point
is a local minimum and therefore stable.

\par
Apart from the quantities whose critical properties are governed by the
exponents given in Eqs.\ (\ref{eqs:3.9}) we are also interested in
the transport properties, and in some other thermodynamic quantities.
Let us first consider the charge diffusion coefficient, $D_{c}$. It
can be obtained from the NL$\sigma$M by a direct calculation of the
particle-hole spin-singlet $q$-$q$ correlation function.\cite{R} To
this end we consider the action $S_{1}$, Eq.\ (\ref{eq:2.7}). If we
replace $Q$ and $\Lambda$ by their mean-field values, Eqs.\ (\ref{eqs:3.1}),
(\ref{eqs:3.2}), we get an action that is quadratic in $q$. Inversion
of the corresponding matrix yields the $q$-$q$ propagator in mean-field
approximation. It has a diffusive structure with diffusion constant
\begin{mathletters}
\label{eqs:3.10}
\begin{equation}
D_{c} = \frac{N_{n=0}^{(0)}}{GH + GK_{s}N_{n=0}^{(0)}}\quad,
\label{eq:3.10a}
\end{equation}
that is, $D_{c}$ vanishes like the OP. The same argument applied to
the particle-hole spin-triplet $q$-$q$ propagator and to a $q$-$q$
correlation that is off-diagonal in replica space, which contain the
spin and heat diffusion coefficients $D_{s}$ and $D_{h}$,
respectively, \cite{R} establish that these transport coefficients
also vanish like the OP,
\begin{equation}
D_{s} = \frac{N_{n=0}^{(0)}}{GH + GK_{t}N_{n=0}^{(0)}}\quad,
\label{eq:3.10b}
\end{equation}
\begin{equation}
D_{h} = N_{n=0}/GH\quad.
\label{eq:3.10c}
\end{equation}
\end{mathletters}%

\par
Finally, we note that the SP approximation is not in any way related to
a systematic small disorder ($G$) expansion. As a consequence, it is
in general not possible to relate a $G$-expansion of the SP theory
to previous work on the NL$\sigma$M,\cite{F1,F2}, or to many-body
perturbation theory.\cite{Cetal} An exception is the $G$-expansion
of the OP itself, see the discussion after Eq.\ (\ref{eq:3.3b}), and
Eq.\ (\ref{eq:A.1}).

\subsection{The Mean-Field Transition as a Renormalization Group Fixed Point}
\label{subsec:III.B}

Here we discuss the mean-field solution presented in the last subsection
in terms of a RG fixed point (FP). We will see that the mean-field
critical behavior is exact for $d>d_{c}^{+}$. We will also use
the RG to derive scaling arguments which allow us to determine the
critical behavior of thermodynamic susceptibilities, such as the specific
heat coefficient and the density susceptibility $\partial n/\partial\mu$,
and of the electrical conductivity.

\par
Consider the action given by Eq.\ (\ref{eq:2.13}), or Eq.\ (\ref{eq:3.8}).
Our parameter space is spanned by $\mu = \{c,t,H,K_{s,t},u_{1},u_{2}\}$.
Here $c$, whose bare value is $1/2G$, is the coefficient of the gradient
squared term in Eq.\ (\ref{eq:2.13}), and $t$ is a measure of the distance
from criticality. Its bare value is $t^{(0)}u/2G^2$ with $t^{(0)}$
given by Eq.\ (\ref{eq:3.4}). $u_{1}
=u/4G^{2}$ and $u_{2} = v/4G^{2}$ are the quartic coupling constants, $H$
is the frequency coupling, and $K_{s,t}$ are the interaction coupling
constants. Since, in anticipation of the RG arguments below, we have
dropped the explicit interaction terms from Eq.\ (\ref{eq:2.13}), this
action contains $K_{s,t}$ only implicitly via the matrix f. The explicit
interaction
dependence is easily restored by adding to Eq.\ (\ref{eq:2.13}) the last
term in Eq.\ (\ref{eq:2.1a}) with $\tilde Q$ replaced by $Q$.
The contributions omitted from Eqs.\ (\ref{eqs:2.10}) yield a term
of the same structure. Structurally,
Eq.\ (\ref{eq:2.13}) is just a matrix version of standard $\phi^{4}$-theory.
Consequently, in looking for a mean-field FP, we follow Ref.\
\onlinecite{MaFisher} and \onlinecite{ZJ}. We define the scale dimension
of a length to be $-1$, and require the exponent $\eta$, i.e. the
anomalous dimension of the $Q$-field, to be zero. The exact scale dimension
of $Q$ is then $[Q]=(d-2)/2$. Power counting then yields the scale
dimensions of the various coupling constants,
\begin{mathletters}
\label{eqs:3.11}
\begin{eqnarray}
[c]=0\quad,\quad [t]=2\quad,\quad [u_{1}]=[u_{2}]=4-d\quad,\nonumber\\
\bigl[ HT\bigr]=(d+2)/2\quad,\quad [K_{s,t}T]=2\quad.
\label{eq:3.11a}
\end{eqnarray}
Sometimes the dimension of temperature, $T$, or frequency, $\Omega$, is
chosen to be $[T]=[\Omega]=d$, which results in $[H]=1-d/2$ and
$[K_{s,t}] = 2-d$. This choice lumps the anomalous part of the dynamical
scaling exponent into the scaling behavior of $H$, and is convenient in
considerations near the lower critical dimension $d_{c}^{-}=2$.\cite{R}
Near the upper critical dimension, $d_{c}^{+}$, the scaling dimension
of $H$ or its equivalent is usually chosen to be zero. Here we adopt
this choice. Then
\begin{equation}
[T] = [\Omega] = \tilde z = d/2 + 1\quad.
\label{eq:3.11b}
\end{equation}
Here $\tilde z$ is a 'naive' dynamical scaling exponent. As we will see
below, the effective dynamical exponent $z$ given in Eq.\ (\ref{eq:3.9b})
results from $\tilde z$ if the existence of dangerous irrelevant variables
is taken into account. Equations\ (\ref{eq:3.11a}),(\ref{eq:3.11b}) give
\begin{equation}
[K_{s,t}] = (2-d)/2\quad.
\label{eq:3.11c}
\end{equation}
\end{mathletters}%
With either choice for $[T]$, $[K_{s,t}]<0$ for $d>2$, that is, the
electron-electron interaction is irrelevant. This justifies, a posteriori,
our omitting the interaction terms in Eq.\ (\ref{eq:2.13}). The same
reasoning implies that the quartic coupling constants $u_{1,2}$ are
irrelevant for $d>4$. Terms generated by expanding the action $S_{2}$ in
Eq.\ (\ref{eq:2.12}) to higher than second order in $\psi$ turn out to be
even more irrelevant than $u_{1}$ and $u_{2}$, as we will see below.
We thus have a Gaussian FP $\mu^{*}=(c,0,0,\ldots)$
which superficially
appears to be stable for $d>4$. The relevant parameters are $t$ and
$T$ or $\Omega$, and the correlation length exponent is $\nu = 1/[t] = 1/2$.
$c$ is marginal, and all other coupling constants are irrelevant. This
Gaussian FP obviously corresponds to the saddle point solution discussed
in the previous subsection.

The zero-loop RG analysis presented
above would imply that the mean-field critical
behavior is exact for $d>4$, unless the RG, at some stage in the
renormalization process, generated new terms in the action which are
relevant for $d>4$. From general power counting it follows that the only
such terms one has to worry about are local terms of $O(Q^{2})$ in
Eq.\ (\ref{eq:2.13}). Such terms would be relevant and grow like $b^{2}$.
We will see in the next section that this indeed
happens, because random-field like terms of this type are generated by
the renormalization process. The net result will be that the upper
critical dimension is
$d_{c}^{+}=6$ rather than $d_{c}^{+}=4$, but the mean-field critical
behavior is still exact for $d>d_{c}^{+}=6$. Also, important aspects
of the scaling arguments that follow from the above zero-loop RG
considerations will remain valid. Before we turn to a more sophisticated
RG approach, we therefore discuss scaling.

\par
The RG arguments given above imply that the OP satisfies a homogeneity
relation,
\begin{mathletters}
\label{eqs:3.12}
\begin{equation}
N(t,\Omega,u_{1},u_{2},\ldots) = b^{1-d/2}\ N(tb^{1/\nu},\Omega b^{\tilde z},
                                 u_{1}b^{4-d},u_{2}b^{4-d},\ldots)\quad,
\label{eq:3.12a}
\end{equation}
with $\nu=1/2$ and $\tilde z = d/2+1$. The exponents $\beta$ and $z$ follow
from Eq.\ (\ref{eq:3.12a}) by means of standard arguments.\cite{MaFisher}
The crucial point is that $u\sim u_{1}\sim u_{2}$ is a dangerous irrelevant
variable (DIV): Solving the saddle point equations explicitly yields
$N(t,\Omega=0,u\rightarrow 0)\sim u^{-1/2}$, and $N(t=0,\Omega,u\rightarrow 0)
\sim u^{-1/3}$. Taking this into account, Eq.\ (\ref{eq:3.12a}) yields
$\beta = 1/2$ and $z=3$ in agreement with Eqs.\ (\ref{eqs:3.9}). In order
to drop the $u_{i}$ from Eq.\ (\ref{eq:3.12a}) we have to change the scale
dimension of $N$, $(d-2)/2$, and the exponent $\tilde z$, to their effective
values $1$ and $z$, respectively,
\begin{equation}
N(t,\Omega) = b^{-1}\ N(tb^{1/\nu},\Omega b^{z})\quad.
\label{eq:3.12b}
\end{equation}
\end{mathletters}%

\par
Let us also determine the singular or critical parts $\chi_{sing}$ of
the thermodynamic susceptibilities $\partial n/\partial \mu$, $\gamma$, and
$\chi_{s}$, where $\gamma_V = \lim_{T\rightarrow 0}C_{V}/T$
is the specific heat
coefficent, and $\chi_{s}$ is the spin susceptibility. To be specific, let
us consider $\gamma_V$. The singular part of the free energy, $f_{sing}$,
satisfies the relation
\begin{mathletters}
\label{eqs:3.13}
\begin{equation}
f_{sing}(t,T,u,\ldots) = b^{-(d+\tilde z)}\ f_{sing}(tb^{1/\nu},Tb^{\tilde z},
                   ub^{4-d},\ldots)\quad.
\label{eq:3.13a}
\end{equation}
In the critical region, $f_{sing}\sim uQ^{4}\sim 1/u$, so that $u$ is a
DIV for $f_{sing}$ as well. This implies,
\begin{equation}
f_{sing}(t,T) = b^{-(4+z)}\ f_{sing}(tb^{1/\nu},Tb^{z})\quad.
\label{eq:3.13b}
\end{equation}
\end{mathletters}%
Differentiating twice with respect to $T$ yields a homogeneity relation
and the critical behavior for $\gamma_{V,sing}$.

\par
More generally, we note that all of these thermodynamic susceptibilites
scale like an inverse volume times a time, so their naive scale dimension
is $d-\tilde z$. The DIV changes this to $4-z=1$, so that
the result for $\gamma_{V,sing}$ actually holds for all of the $\chi_{sing}$,
viz.,
\begin{mathletters}
\label{eqs:3.14}
\begin{equation}
\chi_{sing}(t,T) = b^{-(4-z)}\ \chi_{sing}(tb^{1/\nu},Tb^{z})
                   = b^{-1}\ \chi_{sing}(tb^{2},Tb^{3})\quad,
\label{eq:3.14a}
\end{equation}
or,
\begin{equation}
\chi_{sing}(t,T=0)\sim t^{1/2}\quad,\quad \chi_{sing}(t=0,T)\sim
        T^{1/3}\quad,
\label{eq:3.14b}
\end{equation}
\end{mathletters}%
where we have specialized to the mean-field exponents.

\par
We next consider transport properties. The charge, spin, or heat diffusion
coefficients, which we denote collectively by $D$, all scale like a
length squared times a frequency, so that
\begin{equation}
D(t,\Omega) = b^{2-z}\ D(tb^{1/\nu},\Omega b^{z}) = b^{-1}\ D(tb^{2},
\Omega b^{3}) \quad,
\label{eq:3.15}
\end{equation}
so the diffusion coefficients scale like the OP, Eq.\ (\ref{eq:3.12b}),
in agreement with the
explicit result, Eqs.\ (\ref{eqs:3.10}). We are also interested in the scaling
behavior of the electrical conductivity, $\sigma = D_{c}\partial n/\partial
\mu$. The behavior of $\sigma$ depends on whether or not $\partial n/\partial
\mu$ has an analytic background contribution in addition to the critical
contribution given by Eqs.\ (\ref{eqs:3.14}). In general it is hard to see
why $\partial n/\partial\mu$, or any of the thermodynamic susceptibilites,
should not have an anlytic background contribution, although Ref.\
\onlinecite{Letter1} has given some arguments why the background may be
missing in the particular model under consideration. If $\partial n/\
\partial\mu$ has indeed no background term, then
\begin{mathletters}
\label{eqs:3.16}
\begin{equation}
\sigma(t,\Omega) = b^{-2}\ \sigma(tb^{2},\Omega b^{3})\quad.
\label{eq:3.16a}
\end{equation}
The conductivity exponent $s$, defined by $\sigma(t,\Omega =0)\sim t^{s}$,
is
\begin{equation}
s=2\nu=1\quad.
\label{eq:3.16b}
\end{equation}
If $\partial n/\partial\mu$ does have an analytic background, then $\sigma$
will scale like the diffusion coefficient, so that
\begin{equation}
s=\nu = 1/2\quad.
\label{eq:3.16c}
\end{equation}
\end{mathletters}%

\par
We conclude this section by considering the terms of higher than quadratic
order in $\psi$ that were neglected in deriving Eq.\ (\ref{eq:2.13}).
Expanding the $Tr\ \ln$-term in the action $S_{2}$, Eq.\ (\ref{eq:2.10a}),
in powers of $\psi$, and localizing each term both in real space and in
imaginary time, leads to all possible powers of $\psi$. Schematically we
denote these terms by
\begin{equation}
S_{\psi} = \sum_{n=1}^{\infty}\ w_{n}\ \int d{\bf x}\ \psi^{n}({\bf x})
                    \quad,
\label{eq:3.17}
\end{equation}
where the $w_{n}$ are in general matrices.

\par
Let us examine the Gaussian $\psi$-$\psi$ correlation function in momentum
space. For $d>4$ it goes to a constant at small momenta, for the same
reason for which $v$ in Eq.\ (\ref{eq:2.15b}) is finite for $d>4$. If we
follow Ref.\ \onlinecite{ZJ} in defining the scale dimension of a field
via the behavior of its Gaussian propagator, then we obtain,
\begin{equation}
[\psi] = d/2\quad.
\label{eq:3.18}
\end{equation}
We note that other choices for the scale dimension of $\psi$, e.g.
$[\psi] = d-2$ (motivated by $[Q]=(d-2)/2$ and the fact that $\Lambda$
is conjugate to $Q^{2}$), are also possible, but the final result is
independent of this choice.

\par
We next observe that for fixed $d$, the
$w_{n}$ are divergent if $n$ is large enough. Scaling wavenumbers with
the correlation length $\xi$, we have $w_{n}\sim \xi^{2n-d}$ for
$n\ge d/2$. Therefore, for $n\ge d/2,\ w_{n+1}/w_{n}\sim \xi^{2}$.
Since each successive term in Eq.\ (\ref{eq:3.17}) carries one more
power of $\psi$, we obtain for the scale dimensions of the $w_{n}$,
\begin{mathletters}
\label{eqs:3.19}
\begin{equation}
[w_{n}]=-\frac{d}{2}(n-2)\qquad {\rm for}\quad n\le d/2\quad,
\label{eq:3.19a}
\end{equation}
\begin{equation}
[w_{n+1}]=[w_{n}]-d/2\qquad {\rm for}\quad n<d/2\quad,
\label{eq:3.19b}
\end{equation}
and
\begin{equation}
[w_{n+1}]=[w_{n}]-\frac{d-4}{2}\qquad {\rm for}\quad n\ge d/2\quad.
\label{eq:3.19c}
\end{equation}
\end{mathletters}%
The important conclusions from these power counting arguments are that
all of the $w_{n}$ with $n>2$ are RG irrelevant for $d>4$, that for
fixed $d$ they become more irrelevant as $n$ increases, and that they
all become marginal in $d=4$. Anticipating that we will conclude in the
next subsection that $d_{c}^{+}=6$, this implies that we were justified
in neglecting higher than second order terms in $\psi$, provided we work
either in $d\ge d_{c}^{+}$, or perturbatively (in the sense of an
$\epsilon$-expansion) in $d\alt d_{c}^{+}$.

\par
We conclude this section with three final comments. (1) It is obvious
that the nonlocal gradient corrections to Eq.\ (\ref{eq:3.17}) are even
more irrelevant than the leading terms studied. (2) If one chooses
$[\psi]=d-2$ instead of Eq.\ (\ref{eq:3.18}), then the r.h.s. of Eq.\
(\ref{eq:3.19a}) gets replaced by $-(n-2)(d-2)-(d-4)$, and $d/2$ and
$(d-4)/2$ in Eqs.\ (\ref{eq:3.19b},\ {\ref{eq:3.19c}) get replaced by
$d-2$ and $d-4$, respectively. Obviously, all conclusions drawn above
remain valid. (3) The power counting arguments employed so far do not
distinguish between terms of different symmetry at a given order in powers
of the field. The terms neglected in the $\psi$-expansion could lead to
terms with different symmetries than the ones present in our $Q$-action,
Eq.\ (\ref{eq:2.13}), and, because of complications to be discussed in the
next subsection, such terms could in principle modify the universality
classes for the AMT, even though they are irrelevant by power counting.
We will show in Sec.\ \ref{sec:VI} and in Appendix \ref{app:B} that all
possible terms of this kind are generated by the RG anyway, starting
with Eq.\ (\ref{eq:2.13}), and that they do not modify the critical
behavior near $d=6$.

\section{RENORMALIZATION, AND RANDOM FIELD ASPECTS}
\label{sec:IV}

In this section we consider the renormalization of the field theory derived
in Sec.\ \ref{sec:II}. First we show that a Wilson-type RG procedure
generates additional terms in the action. The most interesting ones are
of random-field (RF) type. Structurally they are identical to the terms
that would have resulted if the original, unreplicated action had had terms
like,
\begin{mathletters}
\label{eqs:4.1}
\begin{equation}
S_{RF} = \sum_{n,\sigma}\ \int d{\bf x}\ h_{n}({\bf x})\ \bar\psi_{\sigma,n}
         ({\bf x})\psi_{\sigma,n}({\bf x})\quad,
\label{eq:4.1a}
\end{equation}
with $h_{n}$ a Gaussian random field with a second moment given by,
\begin{equation}
\bigl\{ h_{n}({\bf x})h_{m}({\bf x})\bigr\} = \Theta (nm)\ \frac{\Delta}{4G}
            \ \delta({\bf x}-{\bf y})\quad.
\label{eq:4.1b}
\end{equation}
\end{mathletters}%
Here $\Delta/4G$ is the strength of the random field, witht the factor
$1/4G$ inserted for convenience. Standard arguments \cite{ImryMa} imply
that such an RF term yields $d_{c}^{+}=6$ instead of $d_{c}^{+}=4$. We then
use a $6-\epsilon$ expansion to describe the AMT in $d=6-\epsilon$
dimensions. A stable RG fixed point is found, and the critical exponents
are calculated to $O(\epsilon)$.

\subsection{Perturbation Theory, and the Renormalization Group}
\label{subsec:IV.A}

Let us consider Eq.\ (\ref{eq:2.13}) at criticality, where $\langle\Lambda
\rangle$ and $\langle Q\rangle$ vanish at zero frequency. Consider the
quartic term,
\begin{equation}
S_{4} = -\frac{u}{4G^{2}}\ \int d{\bf x}\ tr\ Q^{4}({\bf x})\quad,
\label{eq:4.2}
\end{equation}
and the momentum-shell decomposition,\cite{WilsonKogut}
\begin{equation}
Q({\bf x}) = Q^{<}({\bf x}) + Q^{>}({\bf x})\quad.
\label{eq:4.3}
\end{equation}
Here $Q^{>}$ has only Fourier components between $b^{-1}$ and unity,
with $b$ the RG length scale factor, and
$Q^{<}$ has only Fourier components between zero and $b^{-1}$, where we
have set an ultraviolet momentum cutoff equal to unity. Inserting Eq.\
(\ref{eq:4.3}) into Eq.\ (\ref{eq:4.2}), examining terms $\sim (Q^{<})^{2}$,
and averaging over the $Q^{>}$ degrees of freedom, yields to $O(u)$
a term $S_{2,2}$,
\begin{equation}
S_{2,2} = \frac{u}{8G}\ \int_{\bf k,>}\ \frac{1}{k^{2}}\
          \int d{\bf x}\ \sum_{i=\pm}\ \bigl(tr_{i}\ Q^{<}({\bf x})
          \bigr)^{2}\quad.
\label{eq:4.4}
\end{equation}
Here $\int_{\bf k,>}$ denotes an integral over the region $b^{-1}\le k\le 1$.
Examining the structure of $S_{2,2}$ we see that it is indeed a RF like term,
generated by renormalization at one-loop order.

\par
Similarly, if we consider terms of $O(u^{2})$, we find that renormalization
generates new types of quartic terms such as
$(tr_{+}Q^{2})(tr_{-}Q^{2})$ and $(tr_{+}Q^{2})^{2}$. Physically, these
new contributions reflect the fact that in general {\it all} of the terms
in the Landau theory of the AMT are random. For example, a random
quadratic term leads to $(tr_{+}Q^{2})^{2}$. All of these quartic terms
are RG irrelevant for $d>4$. However, one of the main technical points
about RF problems is that the coupling constant, $\Delta$, of a term
like $S_{2,2}$, Eq.\ (\ref{eq:4.4}), scales like $b^{2}$ near the Gaussian
FP. Since the quartic coupling constants, $u_{i}$, scale like $b^{4-d}$
near this FP, products $\Delta u_{i}$ of a RF coupling constant and a
quartic coupling constant scale like $b^{6-d}$, i.e. the product is
marginal in $d=6$. This leads to $d_{c}^{+}=6$.\cite{ImryMa} In Appendix
\ref{app:B} we show that, at least to one-loop order, of all possible
quartic terms only the one that appears in the original action, i.e.
$u\,tr\,Q^{4}$, couples to the RF term. This simplifies the problem
substantially. It means that a term with the structure of $S_{2,2}$,
Eq.\ (\ref{eq:4.4}), is the only one that has to be added to
Eq.\ (\ref{eq:2.13}) in order to perform a one-loop RG analysis.
Expanding $Q$ about its average value, Eq.\ (\ref{eq:3.7}), we can write
the effective action as,
\begin{mathletters}
\label{eqs:4.5}
\begin{eqnarray}
S[\phi] = -\int d{\bf x}\> tr\biggl[\phi ({\bf x})
\Bigl(-\partial_{{\bf x}}^{2}+\langle \Lambda\rangle\Bigr)\phi ({\bf x})
+{u\over G}\Bigl(\bigl(\langle Q\rangle \phi({\bf x})\bigr)^{2}
+\langle Q\rangle^{2}\phi^{2}({\bf x})\Bigr)\biggr]\nonumber\\
+ {\Delta\over 2}\int d{\bf x}\, \sum_{i=>,<}
\Bigl(tr_{i}\, \phi({\bf x})\Bigr)^{2}
- u\int d{\bf x}\> tr\,\phi^{4}({\bf x})\nonumber\\
- {2u\over \sqrt{2G}}\int d{\bf x}\>
tr\,\Bigl[\langle Q\rangle \phi^{3}({\bf x}) +
\phi({\bf x})\langle Q \rangle \phi^{2}({\bf x})\Bigr]
\nonumber\\
- {u\over G}\int d{\bf x}\> tr\,
\biggl[A \Bigl(\phi^{2}({\bf x}) + {2\over \sqrt{2G}}\langle Q\rangle
\phi({\bf x}) \Bigr)\biggr]
- {2\over \sqrt{2G}}\int\! d{\bf x}\> tr\, \Bigl[B\phi ({\bf x})\Bigr] \quad,
\label{eq:4.5a}
\end{eqnarray}
with $A$ and $B$ given by
\begin{equation}
A(\langle Q\rangle,\langle\Lambda\rangle) =
\langle Q\rangle^{2} - 1 + f\bigl(\langle \Lambda
\rangle\bigr)\qquad,\qquad
B(\langle Q\rangle,\langle\Lambda\rangle) =
\langle \Lambda\rangle\langle Q \rangle - 2GH\Omega\qquad,
\label{eq:4.5b}
\end{equation}
\end{mathletters}%
Note that $\langle Q\rangle$ and $\langle\Lambda\rangle$ are determined by
the conditions $\langle\phi\rangle = \langle\psi\rangle = 0$. At zero-loop
order, these conditions yield $A(\langle Q\rangle,\langle\Lambda\rangle)
=B(\langle Q\rangle,\langle\Lambda\rangle)=0$. This is just the zero-loop
equation of state discussed in Sec.\ \ref{sec:III}.

\par
To perform a loop expansion we need the Gaussian (G) propagator for the
field theory defined by Eqs.\ (\ref{eqs:4.5}). In the replica limit
we find,
\begin{mathletters}
\label{eqs:4.6}
\begin{eqnarray}
\langle{^{i}_{r}\phi_{n_{1}n_{2}}^{\alpha_{1}\alpha_{2}}}({\bf k})\
       {^{j}_{s}\phi_{n_{3}n_{4}}^{\alpha_{3}\alpha_{4}}}({\bf p})\rangle
       ^{(G)} = (2\pi)^{d}\ \delta({\bf k}+{\bf p})\ \frac{\delta_{rs}
       \delta_{ij}}{16(k^{2}+m_{12})}\nonumber\\
\times\Bigl[\delta_{13}\delta_{24} + (-)^{r}\delta_{14}\delta_{23} +
       \frac{4\Delta\Theta(n_{1}n_{3})}{k^{2}+m_{12}}\ \delta_{r0}
       \delta_{i0}\delta_{12}\delta_{34}\Bigr]\quad,
\label{eq:4.6a}
\end{eqnarray}
with $1\equiv (n_{1},\alpha_{1})$, etc., and,
\begin{equation}
m_{12} = (\ell_{n_{1}}+\ell_{n_{2}})/2 + u(N_{n_{1}}+N_{n_{2}})^{2}\quad.
\label{eq:4.6b}
\end{equation}
\end{mathletters}%
Note that the last term in Eq.\ (\ref{eq:4.6a}) diverges like $k^{-4}$ at
criticality. In propagator language, this is what changes the upper critical
dimension from $d_{c}^{+}=4$ to $d_{c}^{+}=6$.\cite{Pytteetal} Alternatively,
one can leave the RF term out of the propagator and treat it as an
interaction vertex.\cite{ImryMa} Also note that this anomalously divergent
critical fluctuation is only present in the spin-singlet channel, and only
for correlations that are diagonal in frequency and replica space. This is
in accord with our discussion in Sec.\ \ref{sec:I}.

\subsection{The $\epsilon$-Expansion}

We now follow the standard Wilson-type RG approach to construct differential
recursion relations for the parameters in Eqs.\ (\ref{eqs:4.5}).
Let us first consider the critical theory. Then we can put $\langle Q\rangle
= \langle\Lambda\rangle =0$, and consider only the renormalizations of
$\Delta$ and $u$. The gradient-squared term is not renormalized at one-loop
order, so the field renormalization factor, $\zeta = b^{-(d-2+\eta)/2}$, is
\begin{equation}
\zeta = b^{-(d-2)/2 + O(\epsilon^{2})}\quad,
\label{eq:4.7}
\end{equation}
or, equivalently, the critical exponent $\eta$ is,
\begin{equation}
\eta = O(\epsilon^{2})\quad.
\label{eq:4.8}
\end{equation}
To one-loop order we find the following recursion relations for $u$ and
$\Delta$,
\begin{mathletters}
\label{eqs:4.9}
\begin{equation}
u' = \zeta^{4} b^{d}\ \Bigl[1 - \frac{9}{2}u\Delta\ \int_{\bf k,>}
     \frac{1}{k^{6}}\Bigr]\quad,
\label{eq:4.9a}
\end{equation}
\begin{equation}
\Delta' = \zeta^{2} b^{d}\ \Delta\quad.
\label{eq:4.9b}
\end{equation}
\end{mathletters}%
Taking a differential RG approach, we write
\begin{equation}
\int_{\bf k,>} \frac{1}{k^{6}} = S_{6}\ \ln b + O(\epsilon)\quad,
\label{eq:4.10}
\end{equation}
where $S_{d} = \tilde S_{d}/(2\pi)^{d}$ with $\tilde S_{d}$ the surface of the
$(d-1)$-sphere. Defining
\begin{mathletters}
\label{eqs:4.11}
\begin{equation}
g=u\Delta\quad,
\label{eq:4.11a}
\end{equation}
we obtain the flow equation,
\begin{equation}
\frac{dg}{d\ln b} = \epsilon g - \frac{9}{2}\ S_{6}\ g^{2} + O(g^{3})\quad.
\label{eq:4.11b}
\end{equation}
We see that for $\epsilon >0$, or $d<6$, there is,
besides the unstable Gaussian FP,
a new nontrivial stable FP. The FP value of $g$ is,
\begin{equation}
g^{*} = 2\epsilon/9S_{6} + O(\epsilon^{2})\quad.
\label{eq:4.11c}
\end{equation}
\end{mathletters}

\par
An important feature of Eq.\ (\ref{eq:4.9a}) is that the coefficient does
not depend on whether or not the spin-triplet degrees of freedom ($i=1,2,3$)
are present. This is different from the situation near $d=2$, where the
presence or absence of the triplet channel changes the universality class
of the AMT.\cite{R} Structurally, the insensitivity to the triplet channel
in the present case is due to the fact that the spin-singlet propagator is
more strongly infrared divergent than any other term because it couples to
the random field. Consequently, the leading critical singularities are
controlled by the spin-singlet degrees of freedom.

Next we determine the correlation length exponent, $\nu$. Consider the
disordered or insulator phase, where $N_{n=0}=0$ and $\ell_{n=0}\equiv\ell
\neq 0$. A straightforward renormalization of $\langle\Lambda\rangle$, or
$\ell$, in Eq.\ (\ref{eq:4.5a}), yields,
\begin{equation}
\frac{d\ell}{d\ln b} = 2\ell - gS_6\ell + O(g^{2})\quad.
\label{eq:4.12}
\end{equation}
Similarly, a one-loop renormalization of $A$ and $B$ in
Eq.\ (\ref{eq:4.5a}) gives the equation of state to that order,
\begin{mathletters}
\label{eqs:4.13}
\begin{equation}
N^{2} = 1 - f(\ell) - {Gg \over 4u}\sum_{{\bf p}}{1 \over (p^{2}+\ell)^{2}}
\quad,
\label{eq:4.13a}
\end{equation}
\begin{equation}
\ell N = 2GH\Omega - {gN\over 2}\sum_{{\bf p}}{1 \over (p^{2}+\ell)^{2}}\quad,
\label{eq:4.13b}
\end{equation}
\end{mathletters}%
where both $N$ and $\ell$ should be considered to be functions of $\Omega$.
Using Eq.\ (\ref{eq:4.11c}) in Eq.\ (\ref{eq:4.12}) we find,
\begin{equation}
\ell (b) \sim b^{2[1-\epsilon/9 + O(\epsilon^{2}]}\quad.
\label{eq:4.14}
\end{equation}
In order to determine $\nu$, we finally need the relation between $\ell$ and
the distance from the critical point, $t$. To this end, we expand the r.h.s.
of Eq.\ (\ref{eq:4.13a}) for small $\ell$. The $\ell$-independent
contribution is $t$. At linear order in $d=6$ one finds a term proportional
to $\ell$, and one proportional to $\ell\ln\ell$. The prefactors of these
terms are related by Eq.\ (\ref{eq:2.15a}). Replacing $g$ by $g^{*}$, we
can exponentiate and obtain,
\begin{equation}
\ell \sim t^{1 + \epsilon/18 + O(\epsilon^{2})}\quad.
\label{eq:4.15}
\end{equation}
Combining Eqs.\ (\ref{eq:4.14}) and (\ref{eq:4.15}) gives,
\begin{equation}
t(b) \sim b^{2 - \epsilon/3}\quad,
\label{eq:4.16}
\end{equation}
which identifies the exponent $\nu$ as $1/\nu = 2 - \epsilon/3 + O(\epsilon
^{2})$, or
\begin{equation}
\nu = \frac{1}{2} + \frac{\epsilon}{12}\quad.
\label{eq:4.17}
\end{equation}
Equations (\ref{eq:4.8},\ \ref{eq:4.17}) give $\eta$ and $\nu$ to first order
in $\epsilon$. Standard scaling arguments\cite{MaFisher} yield all other
static exponents. We find,
\begin{eqnarray}
\gamma = 1 + \frac{\epsilon}{6} + O(\epsilon^{2})\quad,\nonumber\\
\beta = \frac{1}{2} - \frac{\epsilon}{6} + O(\epsilon^{2})\quad,\nonumber\\
\delta = 3 + \epsilon + O(\epsilon^{2})\quad.
\label{eq:4.18}
\end{eqnarray}
In order to obtain these results we have used the fact that $u$ is
dangerously irrelevant even for $d<6$, as it is in magnetic RF
problems.\cite{Grinstein} The arguments are the same as those given in
Sec.\ \ref{sec:III} for $d>6$, except that Eqs.\ (\ref{eq:4.9a}) and
(\ref{eqs:4.11}) imply that $u$ scales like
\begin{equation}
u(b) = u(b=0)\ b^{-2}\quad,
\label{eq:4.19}
\end{equation}
instead of $u(b) \sim b^{4-d}$. The net result is that in all $d$-dependent
scaling laws $d$ is replaced by $d-2$. We note that the exponent values
given by Eqs.\ (\ref{eq:4.8},\ \ref{eq:4.17}) and (\ref{eq:4.18}) are
identical with the corresponding ones for the RF Ising model.\cite{ImryMa}
This may not be too surprising. Since the random potential couples only
to the 'longitudinal' field components, $^{0}_{0}Q_{nn}^{\alpha\alpha}$,
the problem is structurally very similar to an anisotropic RF magnetic
model studied by Aharony,\cite{Aharony} which also yielded RF Ising
exponents.

\par
Finally, we need to determine the dynamical exponent $z$. For this purpose
we consider Eq.\ (\ref{eq:4.13b}), which relates $\ell$, $N$, and $\Omega$.
Expanding the r.h.s. for small $\ell$, going to criticality, and
exponentiating yields,
\begin{equation}
N\ \ell^{1 + \epsilon/9 + O(\epsilon^{2})} \sim \Omega\quad.
\label{eq:4.20}
\end{equation}
Using Eq.\ (\ref{eq:4.15}) with $t$ replaced by $N^{1/\beta}$ yields,
\begin{equation}
z = 3 - \frac{\epsilon}{2} + O(\epsilon^{2})\quad.
\label{eq:4.21}
\end{equation}
Note that $z = \delta\beta/\nu = y_{h}$, with $y_{h}$ the exponent of the
field that is conjugate to the OP. This was to be expected, since the RG
did not couple different frequencies, so that, effectively, $\Omega$ in
Eq.\ (\ref{eq:2.1a}) literally acts as the field conjugate to the OP.
The technical reason for this simplification is that the electron-electron
interaction terms turned out to be irrelevant. The physical reason is that
the static RF fluctuations dominate over the quantum fluctucations which
couple different frequencies together.

\section{SCALING DESCRIPTION OF THE ANDERSON-MOTT TRANSITION}
\label{sec:V}

In Sec.\ \ref{sec:III} we gave scaling considerations that applied to
the regime $d>d_c^+$, where mean-field theory gives the correct critical
behavior. Here we generalize this scaling description in the light of
the RG analysis in Sec.\ \ref{sec:IV}.

We first ask which of the general concepts discussed in Sec.\ \ref{sec:III}
will survive. One important general feature is the presence of a dangerous
irrelevant variable (DIV),
$u$.\cite{DIVfootnote} Suppose that $u$ is characterized
by an exponent $\theta$, $u(b)\sim b^{-\theta}$. One-loop perturbation theory
in Sec.\ \ref{sec:IV} gave $\theta = 2+O(\epsilon)$. Although there is
reason to believe that $\theta = 2$ to all orders in the $6-\epsilon$
expansion, this is probably misleading, see the discussion in
Sec.\ \ref{sec:VI}, and we keep $\theta$ general. This adds a third
independent exponent to the usual two independent static exponents.
In addition, there is the dynamical critical exponent $\tilde z$, cf.
Eq.\ (\ref{eq:3.11b}). One effect of the DIV is to effectively change
$\tilde z$ to $z$. In Sec.\ \ref{sec:IV} we found $z$ to be not independent,
but rather to be equal to the exponent $y_h$ of the field conjugate to the
OP. We saw that this was due to the RF fluctuations being stronger than
the quantum fluctuations, and the resulting lack of frequency mixing.
The dominance of RF fluctuations seems to be a general feature of RF
problems. However, it is possible that in low enough dimensions the
explicit interaction terms could become relevant, which would lead to
frequency mixing, and to an independent exponent $z$. For simplicity,
we ignore this possiblity here, and thus write our first scaling law as,
\begin{equation}
z = y_h = \delta\beta/\nu\quad.
\label{eq:5.1}
\end{equation}

The OP obeys a scaling or homogeneity relation,
\begin{mathletters}
\label{eqs:5.2}
\begin{equation}
N(t,\Omega,u,\ldots) = b^{(2-d-\eta)/2}\ N(tb^{1/\nu},\Omega b^{\tilde z},
                          u b^{-\theta},\ldots)\quad,
\label{eq:5.2a}
\end{equation}
which upon elimination of the DIV $u$ turns into
\begin{equation}
N(t,\Omega) = b^{(2+\theta -d-\eta)/2}\ N(tb^{1/\nu},\Omega b^z)\quad.
\label{eq:5.2b}
\end{equation}
\end{mathletters}%
This relates the OP exponent $\beta$ to the three independent exponents
$\nu$, $\eta$, and $\theta$ through the scaling law,
\begin{mathletters}
\begin{equation}
\beta = \frac{\nu}{2}\ (d-\theta -2+\eta)\quad.
\label{eq:5.3a}
\end{equation}
The remaining static exponents are given by the usual scaling laws, with
$d\rightarrow d-\theta$ due to the violation of hyperscaling by the DIV,
\begin{eqnarray}
\delta = (d-\theta + 2-\eta)\nu/2\beta\quad,\nonumber\\
\gamma = \nu(2-\eta)\quad.
\label{eq:5.3b}
\end{eqnarray}
\end{mathletters}%

Now we consider the thermodynamic susceptibilities $\partial n/\partial\mu$,
$\gamma_V$, and $\chi_s$. From the general argument given in connection
with Eqs.\ (\ref{eqs:3.14}) we expect all of them to share the same
critical behavior. Denoting their singular parts collectively by
$\chi_{sing}$ again, the generalization of Eq.\ (\ref{eq:3.14a}) reads,
\begin{equation}
\chi_{sing}(t,T) = b^{-d+\theta +z}\ \chi_{sing}(tb^{1/\nu},Tb^z)\quad.
\label{eq:5.4}
\end{equation}
This links the static critical behavior of the thermodynamic susceptibilities,
characterized by an exponent $\kappa$, $\chi_{sing}(t)\sim t^{\kappa}$,
to that of the OP,
\begin{equation}
\kappa = \beta\quad,
\label{eq:5.5}
\end{equation}
where we have used the scaling laws, Eqs.\ (\ref{eq:5.1},\ \ref{eq:5.3b}).
We see that as a consequence of the dominant RF fluctuations all of the
thermodynamic susceptibilities scale like the OP. This is what we had
found in mean-field theory, Eq.\ (\ref{eq:3.14b}), and we now see that
this is generally valid.

Finally, we consider the transport coefficients. The homogeneity relation
for the diffusion coefficients, Eq.\ (\ref{eq:3.15}), does not contain $d$
explicitly, and therefore is generally valid. If we denote the static
exponent for the diffusion coefficients by $s_D$, $D(t)\sim t^{s_D}$, we
find the following scaling law,
\begin{equation}
s_D = z-2 = \beta - \nu\eta\quad,
\label{eq:5.6}
\end{equation}
The mean-field result that the $D$ scale like the OP is therefore valid
only to the extent that $\eta =0$. The behavior of the electrical
conductivity $\sigma$, which is related to the charge diffusion coefficient by
means of an Einstein relation, $\sigma = D_c\partial n/\partial\mu$,
depends again on whether or not $\partial n/\partial\mu$ has an analytic
background contribution. Generally one would expect such a background
to be present. The conductivity then scales like the charge diffusion
coefficient,
\begin{mathletters}
\label{eqs:5.7}
\begin{equation}
\sigma (t,\Omega) = b^{2-z}\ \sigma(tb^{1/\nu},Tb^z)\quad,
\label{eq:5.7a}
\end{equation}
and the static conductivity exponent $s$ (cf. Eqs.\ (\ref{eqs:3.16})),
is given by
\begin{equation}
s = \frac{\nu}{2}(d-2-\theta-\eta)\quad.
\label{eq:5.7b}
\end{equation}
In models or physical situations where $\partial n/\partial\mu$ has no
analytic background one gets instead,
\begin{equation}
\sigma(t,\Omega) = b^{2-d+\theta}\ \sigma(tb^{1/\nu},\Omega b^z)\quad,
\label{eq:5.7c}
\end{equation}
which leads to
\begin{equation}
s = \nu (d-2-\theta)\quad.
\label{eq:5.7d}
\end{equation}
\end{mathletters}%
In either case, Wegner's scaling law $s=\nu (d-2)$,\cite{Wegner76} which
previously had been found to hold for the AMT\cite{R} as well as the
Anderson transition\cite{LeeRama}, is violated, unless Eq.\ (\ref{eq:5.7b})
holds, and $\theta = 2-d-\eta$. This has profound
consequences which we will discuss in the next section.

\section{DISCUSSION}
\label{sec:VI}

Our results suggest that a crucial physical aspect has been left out of
all previous analyses of the AMT in general, and the NL$\sigma$M commonly
used to describe it in particular. This aspect is the random-field (RF) nature
of the transition, the presence of which has been made plausible in the
Introduction, and which has been confirmed by the explicit RG calculation
in Sec.\ \ref{sec:IV}. In our formulation of the problem the RF aspects
follow naturally from the fact that the random potential couples to the OP
for the AMT, cf. Eq.\ (\ref{eq:1.2}). In this sense an OP description of
the AMT is necessary in order for the RF features to emerge in a
straightforward way, and the absence of an OP theory has been the reason
why these features have not been noted earlier.

The picture of the AMT that emerges from our OP theory with RF aspects
differs crucially in many respects, both physical and technical, from
the one that had previously been obtained by working in the vicinity
of $d=2$.\cite{R} Most importantly, the AMT according to the present
picture is driven by the behavior of the OP, rather than by the soft
modes as in the $2+\epsilon$ expansion. Indeed, we integrated out the
soft modes at an early stage in Sec.\ \ref{sec:II}, and due to the high
dimensionality we are working in they never influence the leading
nonanalytic behavior. Working in a high dimensionality (i.e., in the
vicinity of $d=6$) was in turn forced upon us by the RF aspects of the
theory: It is the dominance of the RF fluctuations over the quantum
fluctuations, which
drive the transition in a $2+\epsilon$ description, that shifts the
upper critical dimension from $d_c^+ =4$ to $d_c^+ =6$.
This is in exact analogy to the case of RF magnets, where the RF fluctuations
dominate over the thermal fluctuations, which is why the RF transition is
sometimes refererred to as a zero-temperature fixed point.\cite{dimred}
As a consequence of the irrelevance of the soft modes we find only one
universality class, irrespective of whether or not the spin-triplet channel
is present. In contrast, the rich variety of universality classes near
$d=2$ resulted from the influence of the diffusive modes in that channel,
the number of which can be controlled by adding magnetic impurities, a
magnetic field, or spin-orbit scattering to the model.\cite{F2,Cetal}
For dimensionalities $d<4$ one expects the soft modes to play an important
role again. In mean-field theory this is readily seen explicitly, see
Appendix \ref{app:A}. However, beyond mean-field theory
the behavior of the model, and even the nature of the
transition, are unclear in that regime, as they are in the magnetic models,
see below.

Another important difference between the present picture and all previous
work on the AMT is that the electron-electron interaction turned out to
be irrelevant, in the RG sense, for all $d>2$. This is another
manifestation of the
subordinate role that quantum fluctuations play in our theory. This leads
to an enormous technical simplification over the work near $d=2$, since
it eliminates the frequency coupling that made the latter extremely
cumbersome.\cite{R} A physical implication of this simplification is that
we find orthodox dynamical scaling, while near $d=2$ there are several
critical time scales, and corresponding dynamical critical exponents, and
one has only what is known as 'weak dynamical scaling'.\cite{fatpaper,R}

Although it is irrelevant for the critical behavior, the electron-electron
interaction is of course necessary for the AMT to exist, since for
noninteracting electrons one has an Anderson transition with an uncritical
DOS. This point is correctly reflected by the theory, as can be seen
from Eq.\ (\ref{eq:2.14}) and has been mentioned after Eqs.\ (\ref{eqs:3.3}).
The electron-electron interaction thus plays a rather trivial, although
crucial, role in the theory: Though irrelevant for the critical
behavior, it ensures that the phase transition under consideration is
accessible. Indeed, for $K_{s,t}\rightarrow 0$ the critical behavior
increases without bound, $G_{c}\rightarrow\infty$. This suggests a
number of distinct phase transition scenarios. The simplest one is that
for sufficiently small interaction constants, or large $G_c$, the AMT
discussed here gets preempted by some other phase transition, such as
a pure Anderson transition. This scenario is particularly likely if
$K_t=0$, and if the electron-electron interaction is short-ranged,
since in this case $K_s$ is irrelevant near the Anderson transition
FP, at least near $d=2$.\cite{R} A different possibility is that in
the above picture the Anderson transition is replaced by an AMT that
is related to the one studied near $d=2$ for the case when either
both $K_s$ and $K_t$ are nonzero, or the electron-electron interaction
is of long range. This picture leaves room for (1) the Anderson transition,
(2) the type of AMT discussed in this paper, and (3) the type of AMT
discussed before.\cite{R} Which transition is actually realized will
depend on the relative strengths of the disorder and the electron-electron
interaction in the various channels.

All thermodynamic susceptibilities, including $\partial n/\partial\mu$,
are found to be critical in the present theory. This is again in
contrast to the results in $d=2+\epsilon$, where $\gamma_V$ and $\chi_s$
may or may not be critical, depending on the universality class, but
$\partial n/\partial\mu$ is {\it never} critical. The reason is that
the critical behavior in $d=2+\epsilon$ derives from logarithmic singularities
in $d=2$, and $\partial n/\partial\mu$, as a frequency integral over a
quantity (the DOS) that is only logarithmically singular itself, cannot
have any logarithmic corrections to any order in perturbation theory.\cite{F1}
We note, however, that on general physical grounds one would actually
expect a nonanalyticity in $\partial n/\partial\mu$, given a nonanalytic
DOS. The frequency integration that takes one from the DOS to
$\partial n/\partial\mu$ may weaken the nonanalyticity, but it cannot
completely remove it. In that respect the result in $d=2+\epsilon$ is
hard to understand, and the present one is physically more plausible.
It also bears some resemblance to the case of a Mott transition in a
Hubbard model, where $\partial n/\partial\mu$ is also
critical.\cite{Hubbardstuff} We emphasize, however,
that for the AMT a critical
$\partial n/\partial\mu$ does not necessarily mean that
$\partial n/\partial\mu$ vanishes at the transition, i.e. that the system
becomes incompressible. As we have mentioned in Secs.\ \ref{sec:III} and
\ref{sec:V}, one expects in general a nonvanishing analytic background
contribution to all thermodynamic quantities except for the OP.
In special models, or for special parameter values, this background
contribution could be absent for some or all of the thermodynamic
susceptibilities, but one would expect a manifest physical reason for
this, like, e.g., a symmetry. Ref.\ \onlinecite{Letter1} argued that
the present model might be such a special case. This suggestion was
based on an explicit representation of the susceptibilities in terms of
the model parameters $H$, $K_s$, and $K_t$,\cite{Cetal2}
and the scaling to zero
under RG iterations of the latter, but no more general physical argument
has been given. Even if the suggestion was correct for this particular
model, however, one would expect more general models to contain noncritical
background contributions. For instance, in Ref.\ \onlinecite{BhattFisher}
it has been argued that local moment effects,
which are absent from our model, lead
to a spin susceptibility that diverges as $T\rightarrow 0$ both in the
metallic and in the insulating phases, but shows no critical behavior at
the MIT. Such noncritical contributions can enter additively to the
critical ones given by Eqs.\ (\ref{eq:5.4}). Of course this would not change
the critical behavior of anything except possibly the conductivity:
Depending on whether or not there is a noncritical
background contribution to $\partial n/\partial \mu$, the critical
exponent $s$ for the conductivity is given by either Eq.\ (\ref{eq:5.7b})
or by Eq.\ (\ref{eq:5.7d}).\cite{backgroundfootnote}
Alternatively, noncritical processes not included
in our model might lead to bare interaction constants in
the action, Eq. (\ref{eq:2.7}), that diverge at $T=0$.
Such terms at most would cause some of the soft modes
in our model to be absent. Because of the irrelevance of the soft modes
mentioned above this would not modify the critical
behavior for $d>4$.

The remarkable analogy which we have found to exist between the AMT and
RF magnetic transitions implies that the AMT will also inherit the
complications that are known to exist for RF magnets, most of which
are not quite understood. For instance, it was mentioned in Sec.\ \ref{sec:V}
that the exponent $\theta$ is probably equal to $2$ to all orders in
perturbation theory, as it is in RF magnets.\cite{theta=2} However, it
is also known that this result is misleading, and that
nonperturbative effects are likely to play an important role in general
RF problems. This is known as the 'dimensional reduction
problem'.\cite{dimred} A possible physical consequence of nonperturbative
effects is the appearance of non-power law dynamical critical behavior, and,
more generally, features of the RF phase transition problem that resemble
those of a glass transition.\cite{FisherVillain} If this is true for the
magnetic RF problem, then one should expect the same for the AMT. A related
question is what happens for $d<4$. Fisher\cite{dimred} has shown that
an infinite number of RF-type operators all become marginal in $d=4$.
Together with the fact that the soft modes will become important for $d<4$
this means that dealing with the problem in $d<4$ will require techniques
substantially different from the ones employed above.

There are also some problems that are germane to the AMT that have
not been addressed yet. In this work we have considered only the
particle-hole degrees of freedom, the particle-particle
or Cooper channel has been omitted. In $d=2+\epsilon$ the Cooper
channel is known to lead to substantial technical problems, which have
not been resolved entirely.\cite{Cooperons} It remains to be seen whether
the Cooper channel is amenable to an easier treatment within the OP
formulation of the AMT. Also, we have restricted ourselves to a short-ranged
model interaction. Our main motivation for this restriction is that a
Coulomb interaction raises questions concerning the treatment of
screening and the plasmon mode which are most naturally addressed within
the context of the general field theory given by Eqs.\ (\ref{eqs:1.1}),
rather than the NL$\sigma$M employed in this paper. Studying the underlying
field theory is also attractive since the main motivation for using the
NL$\sigma$M, viz. its construction as an effective model for the soft modes
in the problem, disappears with the realization that the soft modes are
irrelevant, at least for $d>4$. We leave these problems for future
investigations.

Finally, let us discuss experimental implications of our results.
A result of great importance in this respect is the violation of Wegner
scaling expressed by Eqs.\ (\ref{eqs:5.7}). This removes the requirement
$s\ge 2(d-2)/d$ that followed from Wegner scaling ($s=\nu (d-2)$)
together with the lower bound $\nu\ge 2/d$.\cite{Chayesetal} This
requirement had led to severe problems\cite{R} with the interpretation of
a certain class of experiments on doped semiconductors, most notably
Si:P, which observed $s\approx 1/2$ in $d=3$.
While a dangerous irrelevant variable had been envisaged for some
time as a possible way out of this problem,\cite{DIVSiP} previous
attempts to argue for the existence of one had been found to be
erroneous.\cite{fatpaper}

Another question concerns possible experiments to test whether or not
the AMT has indeed a RF character. In this context an interesting point
is that RF problems are known to contain an anomalously divergent
correlation function.\cite{Grinstein} The point is that in the presence
of a random field the quantity $\{\langle O({\bf x})\rangle\langle O({\bf y})
\rangle\}$, with $O$ the OP, $\langle\ldots\rangle$ the thermodynamic
average, and $\{\ldots\}$ the disorder average, is nonzero even if
$\{\langle O({\bf x})\rangle\}$ vanishes. In the present case the
anomalous correlation takes the form
\begin{mathletters}
\label{eqs:6.1}
\begin{equation}
C({\bf x},{\bf y}) = \{N({\bf x})N({\bf y})\}-N^2\quad,
\label{eq:6.1a}
\end{equation}
with $N({\bf x})$ the unaveraged local DOS at the Fermi level. At the
critical point the Fourier transform of $C$ behaves like
\begin{equation}
C(k\rightarrow 0)\sim k^{-2+\eta-\theta}\quad.
\label{eq:6.1b}
\end{equation}
\end{mathletters}%
The same correlation function, but without the anomalously strong
divergence due to the RF physics, has been considered in connection with
mesoscopic systems,\cite{mesoscopics} and it is measurable, at least in
principle. An experimental observation of the strong divergence predicted
by Eq.\ (\ref{eq:6.1b}) would be a strong indication of the presence of
RF features at the AMT. It would also be interesting to experimentally
look for indications of glassy behavior in the form of
the abovementioned anomalously slow relaxation
processes that one expects in a RF system.

\acknowledgments

We would like to thank John Toner for bringing
Ref.\ \onlinecite{Pytteetal} to our
attention. This work was supported by the NSF under grant numbers
DMR-92-09879 and DMR-92-17496.

\appendix
\section{THE MEAN-FIELD SOLUTION FOR $d<6$}
\label{app:A}

There is no reason to believe that mean-field theory is a particularly good
approximation in the physical dimension $d=3$, or anywhere else below $d=6$.
Here we determine the mean-field results for $d<6$ anyway. We do so
partly for completeness, and partly to illustrate certain interesting
aspects and problems which at the mean-field level are easily soluble
because mean-field theory lacks the complications of the RF aspects of
the transition,
and which may be of relevance for a more complete theory in $d=3$.
Throughout this appendix we suppress the superscript $(0)$ that denoted
the mean-field approximation in Sec.\ \ref{sec:III}.

Before we turn to the critical behavior we consider the frequency dependence
of the OP in the metallic phase.
For small frequencies we find,
\begin{equation}
N(\Omega\rightarrow 0) = N(\Omega =0) + {\rm const}\,\times\,
                           \left\{ \begin{array}{llr}
                           \vert\Omega\vert           & {\rm for} & d>4 \\
                           \vert\Omega\vert^{(d-2)/2} & {\rm for} & 2<d\le 4
                                                     \end{array} \right.\quad.
\label{eq:A.1}
\end{equation}
This low-frequency nonanalyticity, or long-time tail, is the so-called
Coulomb anomaly first found by Altshuler and Aronov \cite{AA} in perturbation
theory. Our mean-field theory correctly reproduces the perturbative result.
In Sec.\ \ref{sec:IV} we have seen that $\Omega$ is the field conjugate to
the OP $N$. This allows for an interesting interpretation of the Coulomb
anomaly: It is the AMT analogue of the fact that in an $O(n)$ ferromagnet
with $n\ge 2$ the magnetic susceptibility diverges everywhere in the
ordered phase.\cite{BW} This is, of course, a consequence of the Goldstone
modes which become increasingly important for $d<4$.

Now we show that the critical behavior is given by Eqs.\ (\ref{eqs:3.9})
for all $d>3$.
We start with the equation of state, Eqs.\ (\ref{eqs:3.2}). It is obvious
that Eq.\ (\ref{eq:3.5}) holds independently of the dimensionality, so
we have $\beta = 1/2$ for all $d$.  If we expand the equation of state
for small frequencies at criticality we obtain for $d>3$,
\begin{equation}
N(\Omega\rightarrow 0) = \vert\Omega\vert^{1/3}\quad.
\label{eq:A.2}
\end{equation}
This determines the exponents $\delta=\beta/\nu z = 3$ for $d>3$.
For $d=3$ there is a logarithmic correction to the $\Omega^{1/3}$, and
for $d<3$ the exponent changes. The fact that the exponent is $d$-independent
for $d>3$, rather than for $d>4$ is a
manifestation of the quantum nature of the mean-field AMT: A quantum
phase transition in $d$ dimensions is expected to be similar to the
corresponding classical transition in $d+1$ dimensions. Note that this
is not the case for the RF phase transition discussed in the main text
because near $d=6$ we found that the interaction terms, which would lead
to frequency mixing, are RG irrelevant.

In order to determine the remaining exponents we turn to Eq.\ (\ref{eq:3.8}).
Obviously $\eta = 0$ for all $d$.
In order to determine $\nu$ we must deal with the divergencies that occur
in the coefficients $u$ and $v$ in $d\le 4$. To protect the divergency,
we keep the momentum dependence of $u$ and $v$, which we then scale with
the correlation length, $\xi$. The integral for $u$ is dominated by the
small frequency region, and we find,
\begin{equation}
u \sim \left\{ \begin{array}{lllr}
               \mbox{const} & & {\rm for} & d\ge 4\\
               \mbox{const} & +\ t^{1/2}\xi^{4-d}      & {\rm for} & 3<d<4
               \end{array} \right.\quad.
\label{eq:A.3}
\end{equation}
The scaling of the correlation length is now obtained by equating the first
and the third term in Eq.\ (\ref{eq:3.8}). We obtain $\xi\sim t^{-1/2}$,
and hence $\nu = 1/2$ for all $d>3$. Again there are logarithmic corrections
to scaling in $d=3$. Equation (\ref{eq:A.3}) immediately gives
$\gamma = 2\nu = 1$, and from $z=y_h=\delta\beta/\nu$ we obtain $z=y_h=3$.
The divergence of $v$ is of no consequence: Since $u$ is effectively
constant for $d>3$, $v$ scales like $\xi^{4-d}\sim t^{-(4-d)/2}$, so
the product $vt$ in Eq.\ (\ref{eq:3.8}) scales to zero.
At the mean-field level, Eqs.\ (\ref{eqs:3.9}) thus hold for $d>3$.

\section{TERMS GENERATED BY THE RENORMALIZATION GROUP}
\label{app:B}

In this appendix we show that the new quartic terms generated by the
RG, which were not in the original action, do not modify the RF fixed
point discussed in Sec.\ \ref{sec:IV}, at least near $d=6$ to one-loop
order. The argument is that these new terms do not couple
to either of the coupling constants $g$ or $\ell$ whose flow equations
are given by Eqs.\ (\ref{eq:4.11b}) and (\ref{eq:4.12}), respectively.

If we repeat the steps below Eq.\ (\ref{eq:4.2}) to $O(u^2)$, then the
RF term in the two-point propagator, Eq.\ (\ref{eq:4.6a}), leads to the
following new quartic terms in the action
\begin{mathletters}
\label{eqs:B.1}
\begin{equation}
S_{4,1} = v_1\ \int d{\bf x}\ tr_+\left(Q^2({\bf x})\right)\
                                         tr_-\left(Q^2({\bf x})\right)\quad,
\label{eq:B.1a}
\end{equation}
\begin{equation}
S_{4,2} = v_2\ \int d{\bf x}\ \left[ tr_+\left(Q^2({\bf x})\right)
                                                        \right]^2\quad,
\label{eq:B.1b}
\end{equation}
\begin{equation}
S_{4,3} = v_3\ \int d{\bf x}\ tr_+\left(Q^2({\bf x})\right)\
                          \left[tr_+\left(Q({\bf x})\right)\right]^2\quad,
\label{eq:B.1c}
\end{equation}
\begin{equation}
S_{4,4} = v_4\ \int d{\bf x} \left[ tr_+\left(Q({\bf x})\right)\right]^4\quad,
\label{eq:B.1d}
\end{equation}
\end{mathletters}%
and additional terms given by $tr_+$ in Eqs.\ (\ref{eqs:B.1}) replaced
by $tr_-$. Note that these quartic terms are distinguished from $S_4$,
Eq.\ (\ref{eq:4.2}), by the appearance of additional traces. Ultimately
these additional traces generate extra replica sums that cause these
terms not to couple into Eqs.\ (\ref{eq:4.11b}) and (\ref{eq:4.12}).

First we argue that these terms cannot contribute to Eq.\ (\ref{eq:4.12}).
We use Eqs.\ (\ref{eq:4.3}) and (\ref{eqs:B.1}), and examine the
$(Q^<)^2$-terms. The RF singularity is necessary to obtain logarithmically
singular contributions in $d=6$, so it is sufficient to use the last
term in Eq.\ (\ref{eq:4.6a}) for the internal propagator,
$\left\langle Q^>Q^>\right\rangle$. This either leads to terms
that vanish in the replica limit, or to terms
proportional to $\left[tr_{\pm}Q\right]^2$.
The latter terms are nonsingular renormalizations of $\Delta$, and can
be neglected near $d=6$.

Next we consider the renormalizations of $O(uv)$ and $O(v^2)$ of $S_4$
or $u$, Eq.\ (\ref{eq:4.2}), where $v$ can stand for any of the coupling
constants in Eqs.\ (\ref{eqs:B.1}).
Let us consider the term of $O(v_2^2)$
specifically, the arguments for all other terms follow the same lines.
In schematic notation, the replica structure of this term is,
\begin{equation}
S_{4,2}^2 \sim Q^{\alpha_1\alpha_2} Q^{\alpha_2\alpha_1}
               Q^{\alpha_3\alpha_4} Q^{\alpha_4\alpha_3}
               Q^{\beta_1\beta_2} Q^{\beta_2\beta_1}
               Q^{\beta_3\beta_4} Q^{\beta_4\beta_3}\quad.
\label{eq:B.2}
\end{equation}
We now use Eq.\ (\ref{eq:4.3}), examine the $(Q^<)^4$-term, and
require that one of the internal propagators be proportional to the
last term in Eq.\ (\ref{eq:4.6a}) in order to obtain a logarithmically
singular term in $d=6$. The resulting replica structure is either
$(tr\ Q^2)^2$, or $(tr\ Q)^2\,tr\ Q^2$. We conclude that this term
does not contribute to the renormalization of $u$.

Similar considerations lead to the general conclusion that none of
the terms given by Eqs.\ (\ref{eqs:B.1}) renormalize $S_4$,
Eq.\ (\ref{eq:4.2}).

\vfill\eject


\begin{references}
\bibitem{Mott} N.F.~Mott, {\it Metal-Insulator Transitions}, Taylor\& Francis
 (London 1990).
\bibitem{R} For a review, see, e.g., D. Belitz and T.~R. Kirkpatrick,
 Rev. Mod. Phys. {\bf 66}, 261 (1994).
\bibitem{F1} A.~M. Finkel'stein, Zh. Eksp. Teor. Fiz. {\bf 84}, 168 (1983)
 [Sov. Phys. JETP {\bf 57}, 97 (1983)].
\bibitem{Wegner79} F. Wegner, Z. Phys. B {\bf 35}, 207 (1979).
\bibitem{F2} A.~M. Finkel'stein, Zh. Eksp. Teor. Fiz. {\bf 86}, 367 (1984)
 [Sov. Phys. JETP {\bf 59}, 212 (1984)].
\bibitem{Cetal} C. Castellani, C. DiCastro, P.~A. Lee, and M. Ma, Phys. Rev.
 B {\bf 30}, 572 (1984).
\bibitem{Hubbardstuff} For a recent review, see, D. Vollhardt in
 {\it Correlated Electron Systems}, edited by V. Emery (World Scientific,
 Singapore, 1993).
\bibitem{MaFisher} See, e.g., S.-K. Ma, {\it Modern Theory of Critical
 Phenomena} (Benjamin, Reading, MA 1976); and M.~E. Fisher, in
 {\it Advanced Course on
 Critical Phenomena}, edited by F.~W. Hahne (Springer, Berlin 1983), p.1
\bibitem{HarrisLubensky} A.~B. Harris and T.~C. Lubensky,
 Phys. Rev. B {\bf 23}, 2640 (1981); see also J.~B. Marston and I. Affleck,
 Nucl. Phys. B {\bf 290[FS20]}, 137 (1987).
\bibitem{OPfootnote} That this really is the most obvious simple OP can be
 seen from Ref.\ \onlinecite{HarrisLubensky}, and from the further development
 in Secs.\ \ref{sec:II}, \ref{sec:III}.
\bibitem{WegnerDOS} F. Wegner, Z. Phys. B {\bf 44}, 9 (1981). Wegner proved
 that the DOS is neither zero nor infinity anywhere in the band for a large
 class of models for noninteracting electrons. This ruled out the theory of
 Ref.\ \onlinecite{HarrisLubensky}, and precludes using the DOS as on OP,
 although nonanalyticities of the DOS within the band are still possible.
\bibitem{FyodorovMirlin} Y.~V. Fyodorov and A.~D. Mirlin, Phys. Rev. Lett.
 {\bf 67}, 2049 (1991).
\bibitem{Letter1} T.~R. Kirkpatrick and D. Belitz, Phys. Rev. Lett. {\bf 73},
 862 (1994).
\bibitem{Letter2} T.~R. Kirkpatrick and D. Belitz, Phys. Rev. Lett. {\bf xx},
 xxx (199x).
\bibitem{ImryMa} Y. Imry and S.~K. Ma, Phys. Rev. Lett. {\bf 35}, 1399 (1975);
 for a review see, e.g., T. Nattermann and J. Villain, Phase Transitions
 {\bf 11}, 5 (1988).
\bibitem{Grinstein} G. Grinstein, Phys. Rev. Lett. {\bf 37}, 944 (1976).
\bibitem{Wegner76} F. Wegner, Z. Phys. B {\bf 25}, 327 (1976).
\bibitem{ZJ} See, e.g., J. Zinn-Justin, {\it Quantum Field Theory and
 Critical Phenomena} (Clarendon, Oxford 1989).
\bibitem{tracefootnote} In our actual RG calculations in Sec.\
 \ref{sec:IV} we ignore the constraint $tr\ Q=0$, Eq.\ (\ref{eq:2.3b}).
 Our rationale is that keeping additional fluctuations cannot possibly
 do any harm, and ignoring this constraint is technically easier than
 enforcing it.
\bibitem{nonpertfootnote} This statement holds perturbatively in an
 $\epsilon$-expansion about $d_{c}^{+}=6$. As in the corresponding magnetic
 RF problems, one expects nonperturbative effects to play a major role which
 is not understood.
\bibitem{AA} B.~L. Altshuler and A.~G. Aronov, Solid State Commun. {\bf 30},
 115 (1979).
\bibitem{WilsonKogut} K.~J. Wilson and J. Kogut, Phys. Rep. {\bf 12}, 75
 (1974).
\bibitem{Pytteetal} E. Pytte, Y. Imry, and D. Mukamel,
 Phys. Rev. Lett. {\bf 46}, 1173 (1981)
\bibitem{Aharony} A. Aharony, Phys. Rev. B {\bf 18}, 3328 (1978).
\bibitem{DIVfootnote} A more careful statement would be that there is
 {\it at least one} DIV. To first order in $\epsilon = 6-d$ there is only
 one DIV, as the one-loop RG analysis presented in Sec.\
 \ref{sec:III} shows. However, work on magnetic RF problems
 suggests that this may change near $d=4$, see Ref.\
 \onlinecite{dimred}. We expect the same conclusion to apply to the AMT.
\bibitem{LeeRama} For a review, see, P.~A. Lee and T.~V. Ramakrishnan, Rev.
 Mod. Phys. {\bf 57}, 287 (1985).
\bibitem{dimred} See, e.g., D.~S. Fisher, Phys. Rev. B {\bf 31}, 7233 (1985),
 and references therein.
\bibitem{fatpaper} T.~R. Kirkpatrick and D. Belitz, Phys. Rev. B {\bf 41},
 11082 (1990).
\bibitem{Cetal2} C. Castellani, C. DiCastro, P.~A. Lee, M. Ma, S. Sorella,
 and E. Tabet, Phys. Rev. B {\bf 33}, 6169 (1986).
\bibitem{BhattFisher} R.~N. Bhatt and D.~S. Fisher, Phys. Rev. Lett.
 {\bf 68}, 3072 (1992).
\bibitem{backgroundfootnote} Questions concerning the scaling of $\sigma$
 and the background contribution to $\partial n/\partial\mu$ could be
 answered by a direct calculation of $\sigma$ as a correlation function.
 Naively, a source term for $\sigma$ will be $~Q^2$, since $\sigma$ is
 related to a two-particle Green function. The mean-field scaling of $Q$
 then suggests $[\sigma] = d-2$. Taking into account the DIV, this is
 consistent with Eqs.\ (\ref{eq:3.16b}) and (\ref{eq:5.7d}). More work
 is needed to confirm or refute this conjecture.
\bibitem{theta=2} A. Aharony, Y. Imry, and S.~K. Ma, Phys. Rev. Lett.
 {\bf 37}, 1364 (1976), G. Parisi and N. Sourlas, Phys. Rev. Lett. {\bf 43},
 744 (1979).
\bibitem{FisherVillain} D.~S. Fisher, Phys. Rev. Lett. {\bf 56}, 416 (1986);
 J. Villain, J. Phys. (Paris) {\bf 46}, 1843 (1985).
\bibitem{Cooperons} T.~R. Kirkpatrick and D. Belitz, Phys. Rev. B {\bf xx},
 xxxx (1994).
\bibitem{Chayesetal} J. Chayes, L. Chayes, D.~S. Fisher, and T. Spencer, Phys.
 Rev. Lett. {\bf 57}, 2999 (1986).
\bibitem{DIVSiP} C. Castellani, G. Kotliar, and P.~A. Lee, Phys. Rev. Lett.
 {\bf 59}, 323 (1987); T.~R. Kirkpatrick and D. Belitz, Phys. Rev. B {\bf 40},
 5227 (1989).
\bibitem{mesoscopics} B.~L. Altshuler, V.~E. Kravtsov, and I.~V Lerner,
 in {\it Mesoscopic Phenomena in Solids}, edited by B.~L. Altshuler,
 P.~A. Lee, and R.~A. Webb, North Holland (Amsterdam 1991).
\bibitem{BW} E. Br$\acute e$zin and D.~J. Wallace, Phys. Rev. B {\bf 7},
 1967 (1973).
\end{references}
\end{document}